\documentclass{article}
\usepackage{graphicx}
\usepackage{verbatim}
\usepackage{amsmath}

\def\permil{\%\raise.10ex\hbox{$_{\scriptstyle 0}$}}

\def\beq{\begin{equation}}
\def\eeq{\end{equation}}
\def\beqn{\begin{eqnarray}}
\def\eeqn{\end{eqnarray}}

\begin{document}

\newcommand{\orders}[1]{\ensuremath{\mathcal{O}\left(s^{#1}\right)}}
\newcommand{\ordershat}[1]{\mathcal{O}\left(\hat{s}^{#1}\right)}
\newcommand{\orderslambda}[1]{\mathcal{O}\left(s_\Lambda^{#1}\right)}
\newcommand{\ordereps}[1]{\ensuremath{\mathcal{O}\left(\epsilon^{#1}\right)}}
\newcommand{\order}[1]{\mathcal{O}\left(#1\right)}
\newcommand{\dk}{d^{D-2}{\bf k}}
\newcommand{\dka}{d^{D-2}{{\bf k}_a}}\newcommand{\dkb}{d^{D-2}{{\bf k}_b}}
\newcommand{\dkl}{d^{D-2}{{\bf k}_l}}\newcommand{\dki}{d^{D-2}{{\bf k}_i}}
\newcommand{\dkone}{d^{D-2}{{\bf k}_1}}
\newcommand{\dktwo}{d^{D-2}{{\bf k}_2}}
\newcommand{\dktwoeps}{\frac{d^{D-2}{{\bf k}_2}}{\mu^{2\epsilon}(2\pi)^{D-4}}}
\newcommand{\dkjet}{d^{D-2}{{\bf k}_J}}
\newcommand{\dkjetone}{d^{D-2}{{\bf k}_{J_1}}}
\newcommand{\dkjettwo}{d^{D-2}{{\bf k}_{J_2}}}
\newcommand{\dkpure}{d^{D-2}{{\bf k}}}
\newcommand{\dkprime}{d^{D-2}{{\bf k}'}}
\newcommand{\dlambdeps}{\frac{d^{D-2}{{\bf \Lambda}}}{\mu^{2\epsilon}(2\pi)^{D-4}}}
\newcommand{\ki}{{\bf k}_i}
\newcommand{\kim}{{\bf k}_{i-1}}
\newcommand{\kip}{{\bf k}_{i+1}}
\newcommand{\kj}{{\bf k}_j}
\newcommand{\kjm}{{\bf k}_{j-1}}\newcommand{\kjp}{{\bf k}_{j+1}}
\newcommand{\km}{{\bf k}_m}
\newcommand{\kmm}{{\bf k}_{m-1}}
\newcommand{\kmp}{{\bf k}_{m+1}}
\newcommand{\kl}{{\bf k}_l}
\newcommand{\klm}{{\bf k}_{l-1}}
\newcommand{\klp}{{\bf k}_{l+1}}
\newcommand{\kn}{{\bf k}_n}
\newcommand{\knm}{{\bf k}_{n-1}}
\newcommand{\knp}{{\bf k}_{n+1}}
\newcommand{\ka}{{\bf k}_a}
\newcommand{\kb}{{\bf k}_b}
\newcommand{\kone}{{\bf k}_1}
\newcommand{\ktwo}{{\bf k}_2}
\newcommand{\kjet}{{\bf k}_J}
\newcommand{\kjetone}{{\bf k}_{J_1}}
\newcommand{\kjettwo}{{\bf k}_{J_2}}
\newcommand{\kpure}{{\bf k}}
\newcommand{\kprime}{{\bf k}'}
\newcommand{\oma}{\omega_0({\bf k}_a)}
\newcommand{\om}[1]{\omega_0(#1)}
\newcommand{\omall}{\omega_0^{LL}({\bf k}_a)}
\newcommand{\omll}[1]{\omega_0^{LL}(#1)}
\newcommand{\del}[1]{\delta^{(2)}\left(#1\right)}
\newcommand{\non}{\nonumber\\}
\newcommand{\asquare}[4]{\left|\mathcal{A}(#1,#2,#3,#4)\right|^2}
\newcommand{\bssquare}[4]{\left|\mathcal{B}_s(#1,#2,#3,#4)\right|^2}
\newcommand{\btssquare}[4]{\left|\widetilde\mathcal{B}_s(#1,#2,#3,#4)\right|^2}
\newcommand{\bsquare}[4]{\left|\mathcal{B}(#1,#2,#3,#4)\right|^2}
\newcommand{\agsquare}[4]{\left|\mathcal{A}_{2g}(#1,#2,#3,#4)\right|^2}
\newcommand{\aqsquare}[4]{\left|\mathcal{A}_{2q}(#1,#2,#3,#4)\right|^2}
\newcommand{\asbar}{\bar\alpha_s}
\newcommand{\seplog}[2]{\ln\frac{s_\Lambda}{\sqrt{#1^2 #2^2}}}
\newcommand{\qi}{{\bf q}_i}\newcommand{\qim}{{\bf q}_{i-1}}\newcommand{\qip}{{\bf q}_{i+1}}
\newcommand{\qj}{{\bf q}_j}\newcommand{\qjm}{{\bf q}_{j-1}}\newcommand{\qjp}{{\bf q}_{j+1}}
\newcommand{\ql}{{\bf q}_l}\newcommand{\qlm}{{\bf q}_{l-1}}\newcommand{\qlp}{{\bf q}_{l+1}}
\newcommand{\qn}{{\bf q}_n}\newcommand{\qnm}{{\bf q}_{n-1}}\newcommand{\qnp}{{\bf q}_{n+1}}\newcommand{\qnpp}{{\bf q}_{n+2}}
\newcommand{\qone}{{\bf q}_1}\newcommand{\qtwo}{{\bf q}_2}
\newcommand{\qa}{{\bf q}_a}\newcommand{\qb}{{\bf q}_b}
\newcommand{\qt}{\tilde{\bf q}}
\newcommand{\dqa}{d^{D-2}{{\bf q}_a}\;}\newcommand{\dqb}{d^{D-2}{{\bf q}_b}\;}
\newcommand{\dqi}{d^{D-2}{{\bf q}_i}\;}
\newcommand{\dqt}{d^{D-2}{\tilde{\bf q}}\;}
\newcommand{\omhat}{\hat{\omega}_0}\newcommand{\omhati}{\hat{\omega}_i}\newcommand{\omhatim}{\hat{\omega}_{i-1}}\newcommand{\omhatl}{\hat{\omega}_l}\newcommand{\omhatn}{\hat{\omega}_n}\newcommand{\omhatlm}{\hat{\omega}_{l-1}}

\newcommand{\shat}{\hat{s}}\newcommand{\that}{\hat{t}}\newcommand{\uhat}{\hat{u}}
\newcommand{\lambd}{{\bf{\Lambda}}}\newcommand{\delt}{{\bf{\Delta}}}

\title{\Large \bf NLO inclusive jet production in $k_T$--factorization}

\author{J.~Bartels$^1$, A.~Sabio~Vera$^2$ and F.~Schwennsen$^1$\\[2.5ex]
$^1$ {\it II. Institut f\"{u}r Theoretische Physik, Universit\"{a}t Hamburg,}\\
 {\it Luruper Chaussee 149, D--22761 Hamburg, Germany}\\
$^2$ {\it Physics Department, Theory Division, CERN,}\\{\it CH--1211, Geneva 23, Switzerland}}

\maketitle

\vspace{-8cm}

\begin{flushright}
{\small CERN--PH--TH/2006--137\\DESY--06--115}
\end{flushright}
\vspace{8.cm}

\begin{abstract}
\noindent
The inclusive production of jets in the central region of rapidity is studied 
in $k_T$--factorization at next--to--leading order (NLO) in QCD 
perturbation theory. Calculations are performed in the Regge limit making use 
of the NLO BFKL results. A jet cone definition is introduced and a 
proper phase--space separation into multi--Regge and 
quasi--multi--Regge kinematic regions is carried out. Two situations are 
discussed: scattering of highly virtual photons, which requires a symmetric 
energy scale to separate the impact factors from the gluon Green's function, 
and hadron--hadron collisions, where a non--symmetric scale choice is needed.
\end{abstract}

\section{Introduction}

The understanding of the physics behind jet production in the context 
of perturbative QCD is an essential ingredient in phenomenological studies 
at present and future colliders. At high energies the theoretical study of 
multijet events becomes an increasingly important task. In the context of 
collinear factorization the calculation of multijet production is complicated 
because of the large number of contributing diagrams. There is, however, a 
region of phase space where it is indeed possible to describe the production 
of a large number of jets: the Regge asymptotics (small--$x$ region) of 
scattering amplitudes. This corresponds to the case where the center--of--mass 
energy in the process under study, $s$, can be considered asymptotically 
larger than any other participating scale. In this limit the dominating 
diagrams are those with gluons being exchanged in the 
$t$--channel. A perturbative analysis of these diagrams shows that it is 
possible to resum contributions of the form $\left(\alpha_s \ln{s}\right)^n$ 
to all orders, with $\alpha_s$ being the coupling constant for the strong 
interaction. This can be achieved by means of the 
Balitsky--Fadin--Kuraev--Lipatov (BFKL) equation~\cite{FKL}.

An essential ingredient in the BFKL approach is the concept of the 
{\it Reggeized gluon} or {\it Reggeon}. In Regge asymptotics colour octet 
exchange can be effectively described by a $t$--channel gluon with its 
propagator being modified by a multiplicative factor depending on a power of 
$s$. This power, also known as {\it gluon Regge trajectory}, depends on the 
transverse momenta of the gluon and is not infrared finite. However, when real 
emissions are included using gauge invariant Reggeon--Reggeon--gluon couplings, 
the divergences cancel out. It is then possible to describe scattering 
amplitudes with any number of particles (jets) in the final state. The 
$\left(\alpha_s \ln{s}\right)^n$ resummation is known as leading--order (LO) 
approximation and provides a simple picture of the underlying physics. 
Nevertheless it is not free of drawbacks, the main two being that, at LO, both 
$\alpha_s$ and the factor scaling the energy $s$ in the resummed logarithms, 
$s_0$, are free parameters not determined by the theory. These limitations can 
be removed if the accuracy in the calculation is increased, and 
next--to--leading (NLO) terms of the form 
$\alpha_s \left(\alpha_s \ln{s}\right)^n$ are taken into 
account~\cite{FLCC}. When this is done, diagrams contributing to the running 
of the coupling have to be included, and also $s_0$ is not longer 
undetermined. As an example, in the context of Mueller--Navelet jets, the 
introduction of NLO effects in the kernel has been recently shown to have 
a large phenomenological impact, in particular, for azimuthal angle decorrelations~\cite{Vera:2006un}.

At LO every Reggeon--Reggeon--gluon vertex corresponds to one single gluon 
emission, and the produced gluon can form a single jet. At NLO the situation is more complicated 
since the emission vertex also contains Reggeon--Reggeon--gluon--gluon and 
Reggeon--Reggeon--quark--antiquark contributions. In the present work we are 
interested in the description of the inclusive production of one jet in the 
BFKL formalism at NLO. This means that the relevant events are those with 
only one jet produced in the central rapidity region of the detector. In order 
to find the probability of production of a single jet it is necessary to 
introduce a jet definition in the emission vertex. This is simple at LO, but 
at NLO we should carefully study the possibility of a double emission in the 
same region of rapidity, leading to the production of one or two jets. This 
will be the main goal of the present paper. Our aim is to clearly separate the different 
contributions to the cross section, and to explain in detail which scales are 
relevant. Particular attention is given to the separation of multi--Regge and 
quasi--multi--Regge kinematics. An earlier analysis has been 
presented in Ref.~\cite{Ostrovsky:1999kj}. We have independently repeated  
these calculations, and we have found several discrepancies which will be 
explained in the text.

Our analysis will be done in two different cases: inclusive jet production in 
the scattering of two photons with large and similar virtualities, and in 
hadron--hadron collisions. In the former case the cross section has a 
factorized form in terms of the photon impact factors and of the  
gluon Green's function which is valid in the Regge limit. In the latter case, 
since the momentum scale of the hadron is substantially lower than the typical 
$k_T$ entering the production vertex, the gluon Green's function for 
hadron--hadron collisions has a slightly different BFKL kernel which, in 
particular, also incorporates some $k_T$--evolution from the nonperturbative, 
and model dependent, proton impact factor to the perturbative jet production 
vertex. We provide analytic formul{\ae} for these two processes, and the 
numerical analysis is left for a future publication.

In the case of hadron--hadron scattering, our cross section formul{\ae} 
contains an {\it unintegrated gluon 
density} which, in addition to the usual dependence on the longitudinal 
momentum fraction typical of collinear factorization, carries an explicit 
dependence on the transverse momentum $k_T$. This scheme is known as 
$k_T$--factorization. So far, no systematic attempt has been made 
to generalize this framework beyond LO accuracy. In the small--$x$ region,
where this type of factorization has attracted particular interest, the 
BFKL framework offers the possibility to formulate, in a systematic way, the 
generalization of the $k_T$--factorization to NLO.  
We therefore interpret our analysis also as a contribution to the 
more general question of how to formulate the unintegrated gluon density and 
the $k_T$--factorization scheme at NLO: our results can be considered as the 
small--$x$ limit of a more general formulation.

After this short introduction, in Section 2 we define, closely following 
Ref.~\cite{Fadin:1998sh}, our notations for the description of a 
general cross section in the BFKL 
approach. We also introduce multi--Regge kinematics (MRK) and the iterative 
structure of the cross sections at LO. In Section 3 we describe the basic 
elements contributing at NLO. The linearity of the BFKL equation remains the 
same while the emission kernel now has several pieces such as 
virtual contributions to one gluon emission and double emissions. 
We describe them in some detail, including a procedure to avoid double counting 
when the MRK is separated from the quasi--multi--Regge kinematics (QMRK). 
The discussion of inclusive jet production starts with a LO description in 
Section 4. Following this introductory part, we present, in Section 5, 
a definition of the NLO jet vertex. We separate the 
different regions of phase space in such a way that the cancellation of 
infrared divergences is explicit for the two cases above--mentioned: 
inclusive jet production in $\gamma^* \gamma^*$ and in hadron--hadron 
interactions. We will also discuss the definition of a NLO unintegrated gluon 
distribution valid in the small--$x$ regime. To close we study in Section 6 
the r{\^ o}le of the scale separating MRK from QMRK and show how, even with 
the jet definition, it is possible to prove that the dependence on this scale 
is power suppressed. Finally, we draw our Conclusions and suggest future 
lines of research.

\section{General structure of BFKL cross sections}

For the sake of clarity, in the present section we introduce the notation 
we will follow in the rest of this study. BFKL cross sections present a 
factorized structure in terms of a universal Green's function, which carries 
the dependence on $s$, and impact factors, which have to be calculated 
for each process of interest. This factorization remains 
unchanged in the transition from LO to NLO. We start by 
defining our normalizations at LO in the following.

Lets consider the case of the total cross section $\sigma_{\rm AB}$ in the 
scattering of two particles A and B. It is convenient to work 
with the Mellin transform
\begin{equation}
  \label{eq:news84}
  {\cal F}(\omega,s_0)=\int_{s_0}^\infty \frac{ds}{s}\left(\frac{s}{s_0}\right)^{-\omega}\sigma_{\rm AB},
\end{equation}
acting on the center--of--mass energy $s$. The dependence on the scaling 
factor $s_0$ belongs to the NLO approximation since the LO calculation is 
formally independent of $s_0$.

If we denote the matrix element for 
the transition ${\rm A} + {\rm B} \rightarrow {\tilde {\rm A}} + {\tilde{\rm  B}} + {\rm n}$ produced particles with momenta $k_i$ ($i=1,\ldots,n$) as 
$\mathcal{A}_{\tilde{\rm A}\tilde{\rm B}+{\rm n}}$, and the corresponding 
element of phase space as $d\Phi_{\tilde{\rm A}\tilde{\rm B}+{\rm n}}$, 
then we can write
\begin{equation}
  \label{eq:news9}
  \sigma_{\rm AB}=\frac{1}{2s}\sum_{n=0}^\infty \int d\Phi_{\tilde{\rm A}\tilde{\rm B}+{\rm n}} |\mathcal{A}_{\tilde{\rm A}\tilde{\rm B}+ {\rm n}}|^2.
\end{equation}
As we mentioned in the Introduction we are interested in the Regge limit 
where $s$ is asymptotically larger than any other scale in the scattering 
process. In this region the 
scattering amplitudes are 
dominated by the production of partons widely separated in rapidity from 
each other. This particular configuration of phase space is known as 
multi--Regge kinematics (MRK). In MRK produced particles are strongly 
ordered in rapidity but there is no ordering of the transverse momenta which 
are only assumed not to be growing with energy.

\begin{figure}[htbp]
  \centering
  \includegraphics[width=8cm]{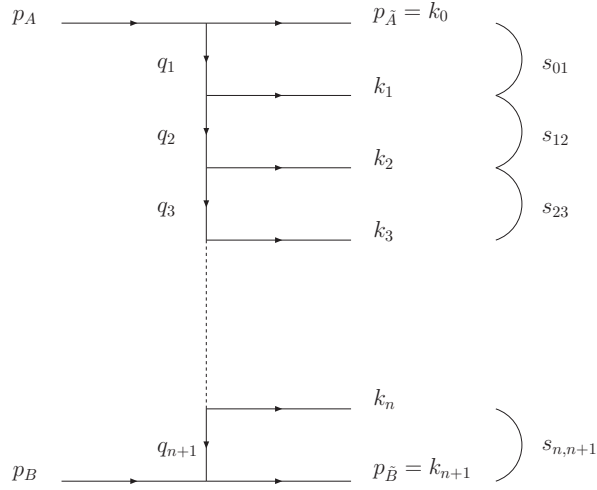}
  \caption{Notation for particle production in MRK.}
  \label{fig:kinematics}
\end{figure}

We fix our notation in Fig.~\ref{fig:kinematics}: $q_i$ correspond to the 
momenta of those particles exchanged in the $t$--channel while the subenergies 
$s_{i-1,i}=(k_{i-1}+k_i)^2$ are related to the rapidity difference between  
consecutive $s$--channel partons. Euclidean two--dimensional transverse 
momenta are denoted in bold. For future discussion we use the Sudakov 
decomposition $k_i = \alpha_i \,p_A + \beta_i \, p_B + {k_i}_\perp$ for the 
momenta of emitted particles.

In MRK the center--of--mass energy for the incoming external particles can be 
expressed in terms of the internal subenergies as
\begin{gather}
  s \simeq\left[\prod_{i=1}^{n+1} s_{i-1,i}\right]\left[\prod_{i=1}^{n} 
\ki^2\right]^{-1}\simeq\sqrt{\qone^2\qnp^2}\,\prod_{i=1}^{n+1} \frac{s_{i-1,i}}{\sqrt{\kim^2\ki^2}},
\label{eq:ssi}
\end{gather}
where we have used the fact that in Regge kinematics $s$ is much larger than 
$-t$ and, therefore, $\alpha_0 \simeq \beta_{n+1} \simeq 1$, 
${\bf k}_0^2 \simeq {\bf q}_1^2 $ and ${\bf k}_{n+1}^2 \simeq 
{\bf q}_{n+1}^2 $. To write down the measure of phase space we use dimensional 
regularization with $D=4 + 2 \, \epsilon$, {\it i.e.}
\begin{gather}
ds\; d\Phi_{\tilde{\rm A}\tilde{\rm B}+{\rm n}} = 2\pi\prod_{i=1}^{n+1}\frac{ds_{i-1,i}}{2 \, s_{i-1,i}}\frac{\dqi}{(2\pi)^{D-1}}.
\label{eq:phasespace}
\end{gather}
The matrix element $\mathcal{A}_{\tilde{\rm A}\tilde{\rm B}+{\rm n}}$ of 
Eq.~(\ref{eq:news9}) can be written in MRK in the factorized form
\begin{equation}
  \label{eq:news17}
  \frac{\mathcal{A}_{\tilde{\rm A}\tilde{\rm B}+{\rm n}}}{2\,s}= \Gamma_A\left[\prod_{i=1}^n \frac{1}{q_i^2}\left(\frac{s_{i-1,i}}{s_{R}}\right)^{\omega_i}\gamma(q_i,q_{i+1})\right]\frac{1}{q_{n+1}^2}\left(\frac{s_{n,n+1}}{s_{R}}\right)^{\omega_{n+1}}\Gamma_B,
\end{equation}
with  $\Gamma_{P}$ being the couplings of the Reggeon to the external 
particles, $\omega_i=\omega(q_i^2)$ the gluon Regge trajectory depending 
on the momentum carried by the Reggeon, and $\gamma(q_i,q_{i+1})$ the 
gauge invariant effective Reggeon--Reggeon--gluon vertices. 
At LO the scale $s_R$ is a free parameter.

Gathering all these elements together it is possible to write the Mellin 
transform of Eq.~(\ref{eq:news84}) as the sum
\begin{eqnarray}
 {\cal F}(\omega,s_0) &=& \sum_{n=0}^\infty {\cal F}^{(n)}(\omega,s_0),
\end{eqnarray}
with the contributions from the emission of n $s$--channel gluons being
\begin{eqnarray}
\frac{{\cal F}^{(n)}(\omega,s_0)}{(2\pi)^{2-D}} = 
\int \left[\prod_{i=1}^{n+1}\dqi\frac{ds_{i-1,i}}{s_{i-1,i}}
\left( \frac{s_{i-1,i}}{s_R}\right) ^{2\omega_i}\left(\frac{s_{i-1,i}}{\sqrt{\kim^2\ki^2}}\right)^{-\omega}\right]&&\nonumber\\
&&\hspace{-9cm}\times \left(\frac{s_0}{\sqrt{\qone^2{\qnp^2}}}\right)^{\omega}\frac{\Phi_A(\qone)}{\qone^2}\left[\prod_{i=1}^n{\cal K}_r(\qi,\qip)\right]\frac{\Phi_B(\qnp)}{\qnp^2} .
\label{eq:lla}
\end{eqnarray}
The impact factors $\Phi_P$ and the real emission kernel for Reggeon--Reggeon into a $s$--channel gluon ${\cal K}_r$ can be written in terms of the square of the vertices $\Gamma_{P}$ and $\gamma$, respectively. The  kernel
${\cal K}_r\left({\bf q}_i,{\bf q}_{i+1}\right)$ 
 is defined such that it includes one gluon propagator on each side:
$\left({\bf q}_i^2 {\bf q}^2_{i+1}\right)^{-1}$. The integration over 
$s_{i-1,i}$ in Eq.~\eqref{eq:lla} takes place from a 
finite $s_0$ to infinity. At LO terms of the form $\omega\ln\ki^2$ or 
$\omega_i \ln s_R$ can be neglected when the integrand is expanded in 
$\alpha_s$. Therefore, at this accuracy, Eq.~\eqref{eq:lla} gives
\begin{eqnarray}
 \frac{{\cal F}^{(n)}(\omega,s_0)}{(2\pi)^{2-D}} =
\int\left[\prod_{i=1}^{n+1}\frac{\dqi}{\omega-2\,\omega_i}\right]\frac{\Phi_A(\qone)}{\qone^{~2}}\left[\prod_{i=1}^n{\cal K}_r(\qi,\qip)\right]\frac{\Phi_B(\qnp)}{\qnp^2},
\label{eq:lla2}
\end{eqnarray}
where the poles in the complex $\omega$--plane correspond to Reggeon 
propagators. This simple structure is a consequence of the linearity of the 
integral equation for the gluon Green's function. We will see below that 
Eq.~\eqref{eq:lla2} holds very similarly at NLO. This fact has been useful in 
the study of different NLO BFKL cross sections using numerical 
techniques in recent years (see Ref.~\cite{MC-GGF}).

After this brief introduction to the structure of BFKL cross 
sections and its iterative expression we now turn to the NLO case. The 
factorization into impact 
factors and Green's function will remain, while the kernel and trajectory 
will be more complex than at LO. We discuss these points 
in the next section.

\section{Different contributions at NLO}

To discuss the various contributions to NLO BFKL cross sections we follow 
Ref.~\cite{Fadin:1998sh}. We comment in more detail those points which 
will turn out to be more relevant for our later discussion of inclusive jet 
production. 
Our starting point are Eqs.~\eqref{eq:news84} to \eqref{eq:phasespace}, which 
remain unchanged. Since at NLO the $s_R$ scale is no longer a free parameter, 
we should modify Eq.~\eqref{eq:news17} to read
\begin{eqnarray}
  \label{eq:news17nlo}
  \frac{\mathcal{A}_{\tilde{\rm A}\tilde{\rm B}+{\rm n}}}{2 \,s} &=& 
\Gamma_A^{(s_{R;0,1})}\left[\prod_{i=1}^n \frac{1}{q_i^2}\left(\frac{s_{i-1,i}}{s_{R;i-1,i}}\right)^{\omega_i}\gamma^{(s_{R;i-1,i},s_{R;i,i+1})}(q_i,q_{i+1})\right]\nonumber\\
&&\hspace{2cm}\times \frac{1}{q_{n+1}^2}\left(\frac{s_{n,n+1}}{s_{R;n,n+1}}\right)^{\omega_{n+1}}\Gamma_B^{(s_{R;n,n+1})}.
\end{eqnarray}
The propagation of a Reggeized gluon with momentum $q_i$ in MRK takes place 
between two emissions with momenta $k_{i-1}$ and $k_i$ 
(see Fig.~\ref{fig:kinematics}). Therefore, at NLO, the term $s_R$, which 
scales the invariant energy $s_{i-1,i}$, does depend on these two consecutive emissions 
and, in general, will be written as $s_{R;i-1,i}$. It is important to note 
that the production amplitudes should be independent of the energy scale 
chosen and, therefore,
\begin{eqnarray}
\Gamma_{A}^{(s_{R;0,1})} = \Gamma_A^{(s_{R;0,1}')} \left(\frac{s_{R;0,1}}{s_{R;0,1}'}\right)^{\frac{\omega_1}{2}},\,\,\,\,\Gamma_{B}^{(s_{R;n,n+1})} = \Gamma_B^{(s_{R;n,n+1}')} \left(\frac{s_{R;n,n+1}}{s_{R;n,n+1}'}\right)^{\frac{\omega_{n+1}}{2}}
\end{eqnarray}
for the particle--particle--Reggeon vertices and
\begin{eqnarray}
\gamma^{(s_{R;i-1,i},s_{R;i,i+1})} \left(q_i,q_{i+1}\right)&=& 
\gamma^{(s_{R;i-1,i}',s_{R;i,i+1}'')} 
\left(q_i,q_{i+1} \right) \nonumber\\
&&\hspace{1cm}\times 
\left(\frac{s_{R;i-1,i}}{s_{R;i-1,i}'}\right)^{\frac{\omega_{i}}{2}}
\left(\frac{s_{R;i,i+1}}{s_{R;i,i+1}''}\right)^{\frac{\omega_{i+1}}{2}}
\end{eqnarray}
for the Reggeon--Reggeon--gluon production vertices.

At NLO, besides the two--loop corrections to the gluon Regge trajectory, 
there are four other contributions which affect the real emission vertex. 
The first one consists of 
virtual corrections to the one gluon production vertex. The second stems 
from the fact that in a chain of emissions widely separated in rapidity 
two of them are allowed to be nearby in this variable, this is known as 
{\it quasi--multi--Regge} kinematics (QMRK). A third source is 
obtained by perturbatively expanding the Reggeon propagators in 
Eq.~\eqref{eq:news17nlo} while keeping 
MRK and every vertex at LO. 
A final fourth contribution is that of the production of 
quark--antiquark pairs. The common feature of all of these new NLO elements 
is that they 
generate an extra power in the coupling constant without building up a corresponding 
logarithm of energy so that $\alpha_s \left(\alpha_s \ln{s}\right)^n$ terms 
are taken into account.

With the idea of introducing a jet definition later on, it is important to 
understand the properties of the production vertex which we now describe 
in some detail. 

Lets start with the virtual corrections to the single--gluon 
emission vertex. These are rather simple and correspond to Eq.~\eqref{eq:lla2} 
with the insertion of a single kernel or impact factor 
with NLO virtual contributions (noted as ($v$)) 
while leaving the rest of the expression 
at Born level (written as ($B$)). More explicitly:
\begin{eqnarray}
\frac{{\cal F}^{(n)}_{\rm virtual}(\omega,s_0)}{(2\pi )^{2-D}}&=&\int\left[\prod_{i=1}^{n+1}\frac{\dqi}{(\omega -2\omega_i)}\right]\nonumber\\
&&\hspace{-3cm}\times\Bigg\{\frac{\Phi^{(B)}_A(\qone)}{\qone^2}\left[\prod_{i=1}^n{\cal K}_r^{(B)}(\qi,\qip)\right]\frac{\Phi^{(v)}_B(\qnp)}{\qnp^2} \nonumber\\
&&\hspace{-2cm} +\frac{\Phi^{(v)}_A(\qone)}{\qone^2}\left[\prod_{i=1}^n{\cal K}_r^{(B)}(\qi,\qip)\right]\frac{\Phi^{(B)}_B(\qnp)}{\qnp^2}\nonumber\\
&&\hspace{-2cm}+\frac{\Phi^{(B)}_A(\qone)}{\qone^2}\sum_{j=1}^n 
\left[\prod_{i=1}^{j-1}{\cal K}_r^{(B)}(\qi,\qip)\right]
{\cal K}_r^{(v)}(\qj,\qjp)\nonumber\\
&&\hspace{1cm}\times\left[\prod_{i=j+1}^{n}{\cal K}_r^{(B)}(\qi,\qip)\right]
\frac{\Phi^{(B)}_B(\qnp)}{\qnp^2}\Bigg\}.
\label{eq:nllavirtual}
\end{eqnarray}

Now we turn to the discussion of how to define QMRK. For this purpose the 
introduction of an extra scale is mandatory 
in order to define a separation in rapidity space between different emissions. 
As in Ref.~\cite{Fadin:1998sh} we call this new scale 
$s_\Lambda$. At LO MRK implies that all $s_{ij}=(k_i+k_j)^2$ are larger than $s_\Lambda$. 
In rapidity space this means that their rapidity difference $|y_i-y_j|$ is larger than 
$\seplog{\ki}{\kj}$. As we stated earlier, in QMRK one single pair of  emissions is allowed to 
be close in rapidity. When any of these two 
emissions is one of the external particles 
$\tilde{\rm A}$ or $\tilde{\rm B}$ it contributes as a real correction to the 
corresponding impact factor. If this is 
not the case it qualifies as a real correction to the kernel. 
This is summarized in the following 
expression where we write real corrections  to the impact factors as ($r$):
\begin{eqnarray}
\frac{{\cal F}^{(n+1)}_{\rm QMRK}(\omega,s_0)}{(2\pi )^{2-D}} &=& 
\int\left[\prod_{i=1}^{n+1}\frac{\dqi}{(\omega -2\omega_i)}\right] \nonumber\\
&&\hspace{-3cm}\times\Bigg\{\frac{\Phi^{(B)}_A(\qone)}{\qone^2}\left[\prod_{i=1}^n{\cal K}_r^{(B)}(\qi,\qip)\right]\frac{\Phi^{(r)}_B(\qnp)}{\qnp^2} \nonumber\\
&&\hspace{-2cm}+\frac{\Phi^{(r)}_A(\qone)}{\qone^2}\left[\prod_{i=1}^n{\cal K}_r^{(B)}(\qi,\qip)\right]\frac{\Phi^{(B)}_B(\qnp)}{\qnp^2}\nonumber\\
&&\hspace{-2cm}+\frac{\Phi^{(B)}_A(\qone)}{\qone^2}\sum_{j=1}^n 
\left[\prod_{i=1}^{j-1}{\cal K}_r^{(B)}(\qi,\qip)\right]{\cal K}_{\rm QMRK}(\qj,\qjp)\nonumber\\
&&\hspace{1cm}\times\left[\prod_{i=j+1}^{n}{\cal K}_r^{(B)}(\qi,\qip)\right]\frac{\Phi^{(B)}_B(\qnp)}{\qnp^2}\Bigg\}.
\label{eq:nllaqmrk}
\end{eqnarray}
The modifications due to QMRK belonging to the kernel or to the impact 
factors are, respectively, ${\cal K}_{\rm QMRK}$ and $\Phi^{(r)}_P$, 
{\it i.e.}
\begin{eqnarray}
{\cal K}_{\rm QMRK}(\qi,\qip) &=& (N_c^2-1)\int d \hat{s} \, \frac {I_{RR} \, \sigma_{RR\rightarrow GG}(\hat{s}) \, \theta(s_{\Lambda}-\hat{s})}{(2\pi)^D \,\qi^2 \, \qip^2}\label{eq:kqmrk},\\
\Phi^{(r)}_P(\kpure) &=& \sqrt{N_c^2-1}\int d \hat{s} \, \frac {I_{PR} \, \sigma_{PR\rightarrow PG}(\hat{s})\, \theta(s_{\Lambda}-\hat{s})}{(2\pi) \, s}
\label{eq:phiqmrk}.
\end{eqnarray}
In both cases $\hat{s}$ denotes the invariant mass of the two emissions in QMRK. The 
Heaviside functions are used to separate the regions of phase space where the emissions 
are at a relative rapidity separation smaller than $s_\Lambda$. It is within 
this region where the LO emission kernel is modified. $\sigma_{RR\rightarrow GG}$ and 
$\sigma_{PR\rightarrow PG}$ are the total cross sections for two Reggeons into two 
gluons, and an external particle and a Reggeon into an external particle and a gluon, 
respectively. $I$ stands for the invariant flux and $N_c$ for the number of colours.

For those sectors remaining in the MRK we use a Heaviside function to keep 
$s_{i-1,i} > s_\Lambda$, in this way MRK is clearly separated from QMRK. We 
then follow the same steps as at LO and use 
Eq.~\eqref{eq:lla} with the modifications already introduced in 
Eq.~\eqref{eq:news17nlo}, {\it i.e.}
\begin{eqnarray}
 \frac{{\cal F}_{\rm MRK}^{(n+1)}(\omega,s_0)}{(2\pi)^{2-D}}&&\nonumber\\
&&\hspace{-2.1cm}=  \int\left[\prod_{i=1}^{n+2}\dqi\frac{ds_{i-1,i}}{s_{i-1,i}}\left( \frac{s_{i-1,i}}{s_{R;i-1,i}}\right) ^{2\omega_i}\left(\frac{s_{i-1,i}}{\sqrt{\kim^2\ki^2}}\right)^{-\omega}\theta(s_{i-1,i}-s_\Lambda)\right]\nonumber\\
&& \hspace{-1.2cm}\times \left(\frac{s_0}{\sqrt{\qone^2\qnpp^2}}\right)^{\omega}\frac{\Phi_A^{(B)}(\qone)}{\qone^2}\left[\prod_{i=1}^{n+1}{\cal K}_r^{(B)}(\qi,\qip)\right]\frac{\Phi_B^{(B)}(\qnpp)}{\qnpp^2} .
\label{eq:nlla}
\end{eqnarray}
After performing the integration over the $s_{i-1,i}$ variables the following 
interesting dependence on $s_\Lambda$ arises:
\begin{eqnarray}
 \frac{{\cal F}_{\rm MRK}^{(n+1)}(\omega,s_0)}{(2\pi )^{2-D}} &=&  \int\left[\prod_{i=1}^{n+2}\frac{\dqi}{(\omega-2\omega_i)}\left( \frac{s_\Lambda}{s_{R;i-1,i}}\right)^{2\omega_i}\left(\frac{s_\Lambda}{\sqrt{\kim^2 \ki^2}}\right)^{-\omega}\right]\nonumber\\
&& \hspace{-2cm}\times \left(\frac{s_0}{\sqrt{\qone^2\qnpp^2}}\right)^{\omega}\frac{\Phi_A^{(B)}(\qone)}{\qone^2}\left[\prod_{i=1}^{n+1}{\cal K}_r^{(B)}(\qi,\qip)\right]\frac{\Phi_B^{(B)}(\qnpp)}{\qnpp^2}.
\end{eqnarray}

It is now convenient to go back to Eq.~(\ref{eq:news84}) and write the lower 
limit $s_0$ of the Mellin transform as a 
generic product of two 
scales related to the external impact factors, {\it i.e.} 
$s_0=\sqrt{s_{0;A} \, s_{0;B}}$. By expanding in $\alpha_s$ the factors 
with powers in $\omega$ and $\omega_i$ it is then possible to identify the 
NLO terms:
\begin{eqnarray}
 \frac{{\cal F}_{\rm MRK}^{(n+1)}(\omega,s_0)}{(2\pi )^{2-D}} =  \int\left[\prod_{i=1}^{n+2}\frac{\dqi}{(\omega-2\omega_i)}\right]\frac{\Phi_A^{(B)}(\qone)}{\qone^2}\left[\prod_{i=1}^{n+1}{\cal K}_r^{(B)}(\qi,\qip)\right]\frac{\Phi_B^{(B)}(\qnpp)}{\qnpp^2}&&\hspace{-0.8cm}\nonumber\\
&& \hspace{-12cm}\times\Bigg\{ 1 -\frac{\omega}{2}\ln\frac{s_\Lambda^2}{\kone^2 s_{0;A}}+\omega_1\ln\frac{s_\Lambda^2}{s_{R;0,1}^2}-\sum_{i=2}^{n+1}\left[\frac{\omega}{2}\ln\frac{s_\Lambda^2}{\kim^2 \ki^2}-\omega_i\ln\frac{s_\Lambda^2}{s_{R;i-1,i}^2}\right]\hspace{-0.8cm}\nonumber\\
&& \hspace{-12cm}\phantom{\times\Bigg[ 1}-\frac{\omega}{2}\ln\frac{s_\Lambda^2}{\knp^2 s_{0;B}}+\omega_{n+2}\ln\frac{s_\Lambda^2}{s_{R;n+1,n+2}^2}\Bigg\}.\hspace{-0.8cm}
\end{eqnarray}
To combine this expression with that of the QMRK contribution we should make a choice 
for $s_R$. The most convenient one is $s_{R;i,j}=\sqrt{s_{R;i} \, s_{R;j}}$, where  
for intermediate Reggeon propagation we use $s_{R;i}=\ki^2$, and for the connection with the 
external particles $s_{R;0}= s_{0;A}$ and $s_{R;n+2}=s_{0;B}$. We can then 
write
\begin{eqnarray}
 \frac{{\cal F}_{\rm MRK}^{(n+1)}(\omega,s_0)}{(2\pi )^{2-D}} =  \int\left[\prod_{i=1}^{n+2}\frac{\dqi}{(\omega-2\omega_i)}\right]\frac{\Phi_A^{(B)}(\qone)}{\qone^2}\left[\prod_{i=1}^{n+1}{\cal K}_r^{(B)}(\qi,\qip)\right]\frac{\Phi_B^{(B)}(\qnpp)}{\qnpp^2}&& \hspace{-0.9cm}\nonumber\\
&& \hspace{-12cm}\times\Bigg\{ 1 -\frac{(\omega-2\omega_1)}{2}\ln\frac{s_\Lambda^2}{\kone^2 s_{0;A}}-\sum_{i=2}^{n+1}\left[\frac{(\omega-2\omega_i)}{2}\ln\frac{s_\Lambda^2}{\kim^2 \ki^2}\right]\nonumber\\
&& \hspace{-7cm}-\frac{(\omega-2\omega_{n+2})}{2}\ln\frac{s_\Lambda^2}{\knp^2 s_{0;B}}\Bigg\}.
\label{MRKfinal}
\end{eqnarray}
This corresponds to the LO result for ${\cal F}^{(n+1)}$ plus additional 
terms where the $\omega-2\omega_i$ factor cancels, in such a way that they 
can be combined with the LO result of ${\cal F}^{(n)}$. 

The quark contribution can be included in a straightforward manner since 
between the quark--antiquark emissions there is no propagation of a Reggeized 
gluon. In this way one can simply write
\begin{multline}
\frac{{\cal F}^{(n+1)}_{Q\bar{Q}}(\omega,s_0)}{(2\pi )^{2-D}}=\int\left[\prod_{i=1}^{n+1}\frac{\dqi}{(\omega -2\omega_i)}\right]\frac{\Phi^{(B)}_A(\qone)}{\qone^2}\frac{\Phi^{(B)}_B(\qnp)}{\qnp^2}\\
\times\sum_{j=1}^n \left[\prod_{i=1}^{j-1}{\cal K}_r^{(B)}(\qi,\qip)\right]{\cal K}_{Q\bar{Q}}(\qj,\qjp)\left[\prod_{i=j+1}^{n}{\cal K}_r^{(B)}(\qi,\qip)\right].
\label{eq:nllaqqbar}
\end{multline}
The production kernel can be written as
\begin{align}
{\cal K}_{Q\bar{Q}}(\qi,\qip) ~=&~ (N_c^2-1)\int d \hat{s} \, \frac {I_{RR} \, 
\sigma_{RR\rightarrow Q\bar{Q}}(\hat{s})}{(2\pi)^D \, \qi^2 \, \qip^2},
\label{eq:krrqq}
\end{align}
with $\sigma_{RR\rightarrow Q\bar{Q}}$ being the total cross section for two Reggeons 
producing the quark--antiquark pair with an invariant mass ${\hat s}$.

The combination of all the NLO contributions together generates the 
following expression for the NLO cross section: 
\begin{eqnarray}
{\cal F}(\omega ,s_0)_{AB}&=&\sum_{n=0}^{\infty}\frac{1}{(2\pi )^{D-2}}\int\left[\prod_{i=1}^{n+1}\frac{\dqi}{(\omega -2\omega_i)}\right]\nonumber\\
&&\hspace{0.4cm}\times\frac{\Phi_A(\qone;s_{0;A})}{\qone^2}\left[\prod_{i=1}^n{\cal K}_{r}(\qi,\qip)\right]\frac{\Phi_B(\qnp;s_{0;B})}{\qnp^2},
\end{eqnarray}
where the NLO real emission kernel contains several terms:
\begin{eqnarray}
{\cal K}_{r}(\qi,\qip) &=& 
\left( {\cal K}_r^{(B)} + {\cal K}_r^{(NLO)} \right) (\qi,\qip) \nonumber\\
&=& \left( {\cal K}_r^{(B)} + {\cal K}_r^{(v)} + 
{\cal K}_{GG} + {\cal K}_{Q\bar Q} \right) (\qi,\qip)
\label{eq:kernelnlo},
\end{eqnarray}
with ${\cal K}_{Q\bar Q}$ given by Eq.~\eqref{eq:krrqq}. The two gluon 
production kernel ${\cal K}_{GG}$ is the combination of ${\cal K}_{\rm QMRK}$ of Eq.~\eqref{eq:kqmrk} 
and the MRK contribution in Eq.~(\ref{MRKfinal}). It explicitly reads
\begin{eqnarray}
{\cal K}_{GG} (\qi,\qip) &=& (N_c^2-1)\int d \hat{s} \frac {I_{RR} \sigma_{RR\rightarrow GG}(\hat{s}) \, \theta(s_{\Lambda}-\hat{s})}{(2\pi)^D \, \qi^2 \, \qip^2}\nonumber\\
&&\hspace{-3cm}- \int\dqt \, {\cal K}_r^{(B)}(\qi,\qt) \, {\cal K}_r^{(B)}(\qt,\qip)\frac{1}{2}\ln \left(\frac{s_{\Lambda}^2}{(\qi-\qt)^2(\qip-\qt)^2}\right). 
\label{eq:krrgg}
\end{eqnarray}
Below we will show that when $s_\Lambda$ is taken to infinity the second term 
of this expression subtracts the logarithmic divergence of the first one. 
When computing the total cross section it is natural to remove the dependence 
on the parameter $s_\Lambda$ in this way. For our jet production cross 
section, however, we prefer to retain the dependence upon $s_\Lambda$.

For the impact factors a similar expression including virtual and MRK corrections as in 
Eq.~\eqref{eq:phiqmrk} arises:
\begin{eqnarray}
\label{eq:impactfactornlo}
\Phi_P(\qone;s_{0;P}) &=& \Phi_P^{(B)}+\Phi_P^{(v)}+\sqrt{N_c^2-1}\int d \hat{s} \frac{I_{PR} \, \sigma_{PR}(\hat{s}) \, \theta(s_{\Lambda}-\hat{s})}{(2\pi) \, s}\nonumber\\
&&\hspace{-1cm}-\int \dqt \, \Phi^{(B)}_P(\qt) \, {\cal K}_r^{(B)}(\qt,\qone)\frac{1}{2}\ln \left(\frac{s_{\Lambda}^2}{(\qone-\qt)^2s_{0;P}}\right).
\end{eqnarray}
From this expression it is now clear why to choose the factorized form 
$s_0=\sqrt{s_{0;A} \, s_{0;B}}$: in this way each of the impact factors $\Phi_{A,B}$ 
carry its own $s_{0;A,B}$ term at NLO independently of the choice of scale in the other.

To conclude this section, for the sake of clarity, the different contributions to 
the NLO BFKL kernel

\begin{center}
\begin{figure}[htbp]
  \centering
  \includegraphics[width=12cm]{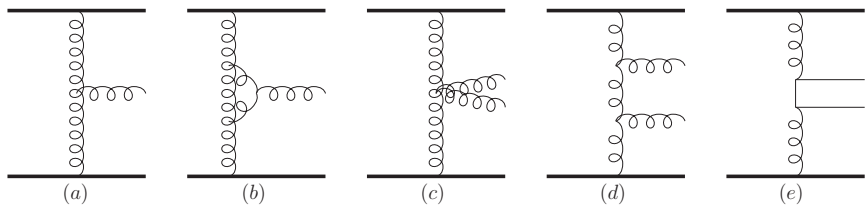}
  \caption{Contributions to real emission kernel at LO (a) and NLO (b-e).}
  \label{fig:grouping}
\end{figure}
\end{center}

\begin{center}
\begin{tabular}[h]{lcc}
CONTRIBUTION            & NUMBER OF EMISSIONS& Fig.\ref{fig:grouping}\\
\hline
MRK @ LO             & $n$                      & (a)\\
\hline
Virtual                 & $n$                      & (b)\\
QMRK                    & $n+1$                    & (c)\\
MRK @ NLO            & $n+1$                    & (d)\\
Quark--antiquark pair    & $n+1$                    & (e)\\
\hline
\end{tabular}
\end{center}
are pictorially represented in Fig.~\ref{fig:grouping}.
As a final remark we would like to indicate that the divergences present in 
the gluon trajectories $\omega_i$ (see Ref.~\cite{FLCC}) are all cancelled 
inside the inclusive terms. We will see how the soft and collinear 
divergences of the production vertex are either cancelled amongst its 
different components or are 
regularized by the jet definition.

After having introduced the notation and highlighted the different 
constituents of a BFKL production kernel at NLO, in the coming section we 
describe how to calculate the inclusive production of jets in two different 
environments. The first one is the case of the interaction between two 
small and perturbative objects, highly virtual photons, and 
the second will be the collision of two large and non--perturbative 
external particles such as the ones taking place at hadron--hadron colliders.

\section{Inclusive jet production at LO}

As MRK relies on the transverse scales of the emissions and internal lines 
being of the same order it is natural to think that processes characterized 
by two large and similar transverse momenta are the ideal environment for 
BFKL dynamics to  show up. Moreover, as the resummation is based on 
perturbative degrees of freedom, these large scales associated to the external 
particles should favor the accuracy of the predictions. 
An ideal scenario is the interaction between two photons with large 
virtualities $Q_{1,2}^2$ in the Regge limit $s \gg |t|\sim Q_1^2\sim Q_2^2$. 
The total cross section for this process has been investigated in a large 
number of papers in recent years. Here we are interested in the inclusive 
production of a single jet in the central region of rapidity in this process. 
We will consider the case where the transverse momentum of the jet is of the 
same order as the virtualities of the photons.

As a starting point we review single jet production at LO accuracy. As usual 
the total cross section can be written as a convolution of the photon impact 
factors with the gluon Green's function, {\it i.e.}
\begin{equation}
  \label{eq:total}
\sigma(s) = \int\frac{d^2 {\bf k}_a}{2\pi\ka^2}\int\frac{d^2{\bf k}_b}{2\pi\kb^2} \, 
\Phi_A(\ka) \, \Phi_B(\kb) \,\int_{\delta-i\infty}^{\delta+i\infty}\frac{d\omega}{2\pi i} \left(\frac{s}{s_0}\right)^\omega f_\omega(\ka,\kb).
\end{equation}
A  common choice for the energy scale is $s_0=|\ka|\,|\kb|$ which naturally 
introduces the rapidities $y_{\tilde A}$ and $y_{\tilde B}$ of the emitted particles with momenta $p_{\tilde A}$ and $p_{\tilde B}$ since
\begin{equation}
  \left(\frac{s}{s_0}\right)^\omega = e^{\omega(y_{\tilde A}-y_{\tilde B})}.
\end{equation}
Let us remark that a change in this scale can be treated as a redefinition
of the impact factors and, if $s_0$ is chosen to depend only on $\ka$ or
only on $\kb$, the kernel as well. This treatment lies beyond LO and will be discussed in the next section. 
The gluon Green's function $f_\omega$ corresponds to the solution of the BFKL 
equation
\begin{gather}
  \label{eq:bfklequation}
  \omega f_\omega(\ka,\kb) = \delta^2(\ka-\kb)+\int d^2{\bf k}\;\mathcal{K}(\ka,\kpure)f_\omega(\kpure,\kb),\\
  \mathcal{K}(\ka,\kpure) = 2 \, \omega(\ka^2) \, 
\delta^{2}(\ka-\kpure) + \mathcal{K}_r(\ka,\kpure),
\end{gather}
where the kernel ${\cal K}$ contains a term related to the Reggeized gluon 
propagator, the trajectory $\omega (\ka^2)$, and the real emission kernel, 
${\cal K}_r$.

\begin{center}
\begin{figure}[htbp]
  \centering
  \includegraphics[height=7cm]{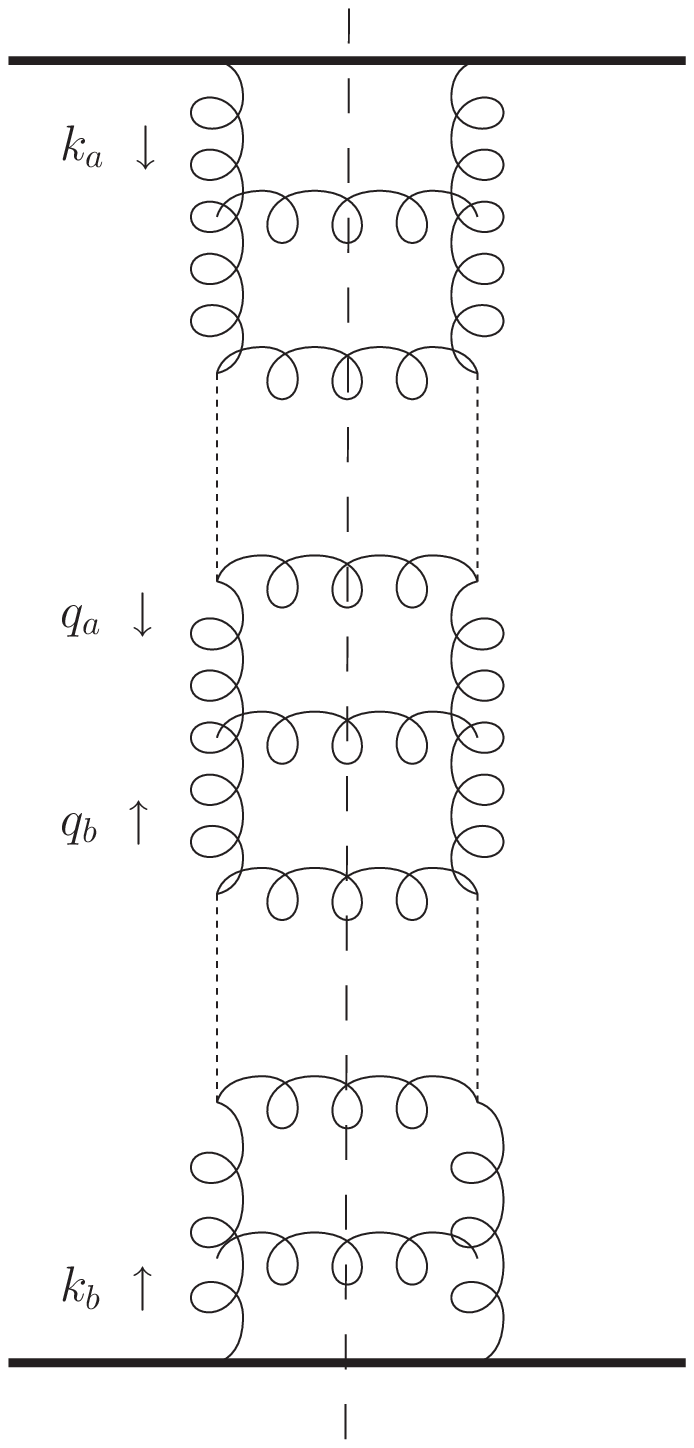}
  \hspace{1cm}
  \includegraphics[height=7cm]{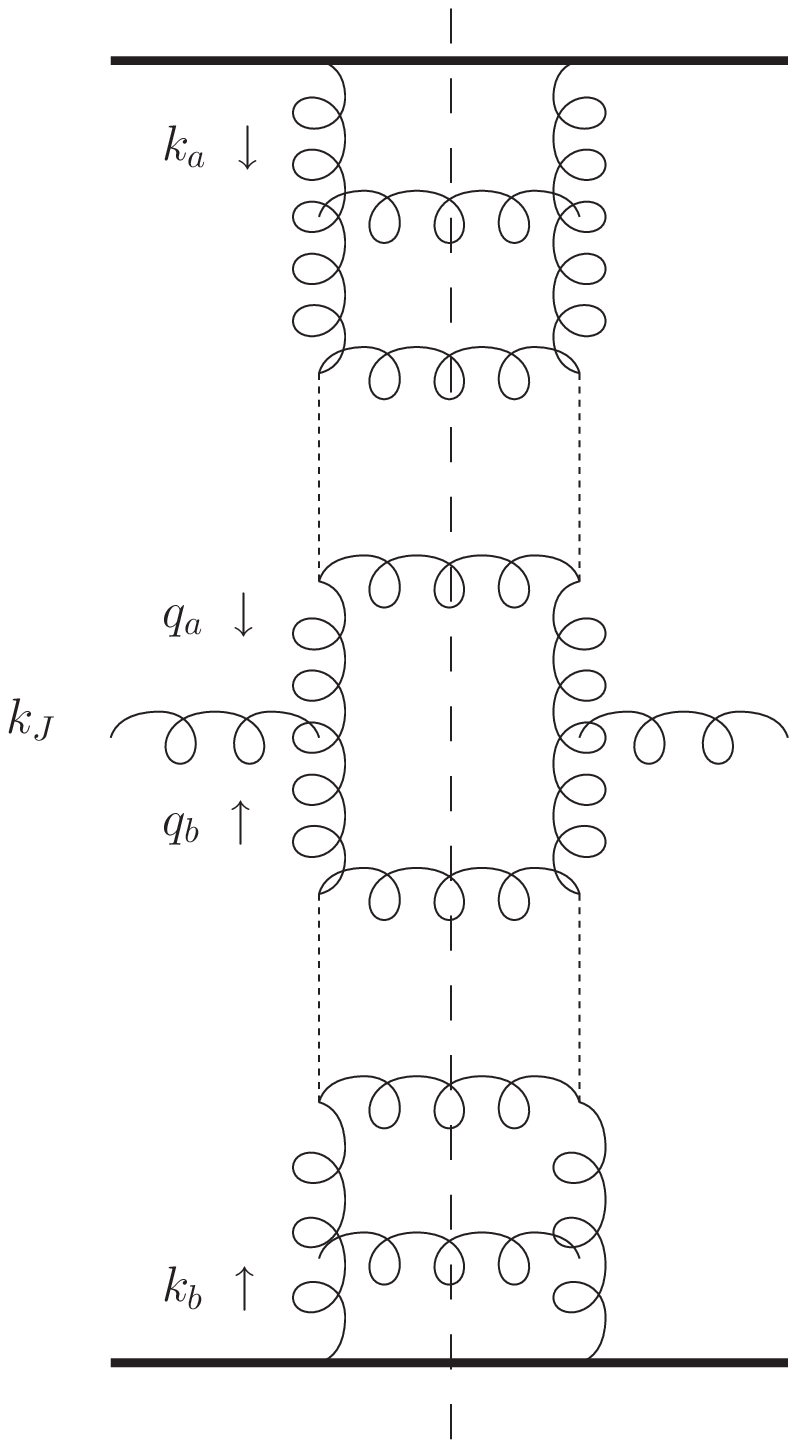}
  \caption{Total cross section and inclusive one jet production in the BFKL 
approach.}
  \label{fig:crosslo}
\end{figure}
\end{center}

For the inclusive production of a single jet we assign to it a 
rapidity $y_J$ and a transverse momentum $\kjet$, as shown in 
Fig.~\ref{fig:crosslo}. In this way, if 
$k_J=\alpha_J p_A+\beta_J p_B+k_{J \perp}$ the corresponding rapidity is 
$y_J=\frac{1}{2}\ln\frac{\alpha_J}{\beta_J}$. Using its on--shell condition 
we can write 
\begin{equation}
   k_J=\sqrt{\frac{\kjet^2}{s}}e^{y_J}p_A+\sqrt{\frac{\kjet^2}{s}}
e^{-y_J}p_B+k_{J \perp}.
\end{equation}

It is possible to single out one gluon emission by extracting its  
emission probability from the BFKL kernel. 
The differential cross section in terms of the jet variables can then be 
constructed in the following way:
\begin{multline}
\frac{d\sigma}{d^{2}\kjet dy_J} =
\int\frac{d^2 \ka}{2\pi\ka^2}\int\frac{d^2 \kb}{2\pi\kb^2} \, \Phi_A(\ka) 
\, \Phi_B(\kb)\\
\times\int d^2 \qa \int d^2 \qb \int_{\delta-i\infty}^{\delta+i\infty}
\frac{d\omega}{2\pi i} \left(\frac{s_{AJ}}{s_0}\right)^\omega 
f_\omega(\ka,\qa)\\
\times  
{\cal V}(\qa,\qb;\kjet,y_J)\int_{\delta-i\infty}^{\delta+i\infty}\frac{d\omega'}{2\pi i} \left(\frac{s_{BJ}}{s_0'}\right)^{\omega'}f_{\omega'}(-\qb,-\kb)
\label{eq:masterformula0}
\end{multline}
with the LO emission vertex being
\begin{equation}
{\cal V}(\qa,\qb;\kjet,y_J) = \mathcal{K}_r^{(B)}\left(\qa,-\qb\right) \, \del{\qa+\qb-\kjet}. 
\label{eq:jetvertexloprelim}
\end{equation}
By selecting one emission to be exclusive we have factorized the gluon Green's 
function into two components. Each of them connects one of the external 
particles to the jet vertex. In the notation of Eq.~\eqref{eq:masterformula0} 
the energies of these blocks are
  \begin{align}
    s_{AJ} =& (p_A+q_b)^2, & s_{BJ} =& (p_B+q_a)^2 .
  \end{align}
In a symmetric situation, where the jet provides a hard scale as well as the impact factors, a natural choice for the scales is similar to that in the total cross section
\begin{align}
  s_0 =& |\ka|\,|\kjet|, & s_0' =& |\kjet|\,|\kb|.
\label{sosoprime}
\end{align}
These choices can now be related to the relative rapidity between the jet and the external particles. To set the ground for the NLO discussion of the next section we introduce an additional integration over the rapidity $\eta$ of the central system:
\begin{multline}
\frac{d\sigma}{d^{2}\kjet dy_J} =
\int d^2 \qa \int d^2 \qb \int d\eta\\
\times\left[\int\frac{d^2 \ka}{2\pi\ka^2} \, \Phi_A(\ka) \, 
\int_{\delta-i\infty}^{\delta+i\infty}\frac{d\omega}{2\pi i} e^{\omega(y_A-\eta)} f_\omega(\ka,\qa)\right]\, \mathcal{V}(\qa,\qb,\eta;\kjet,y_J)\\
\times \left[\int\frac{d^2 \kb}{2\pi\kb^2} \, 
\Phi_B(\kb) \int_{\delta-i\infty}^{\delta+i\infty}\frac{d\omega'}{2\pi i} e^{\omega'(\eta-y_B)}f_{\omega'}(-\qb,-\kb)\right] 
\label{eq:masterformula1}
\end{multline}
with the LO emission vertex being
\begin{equation}
\mathcal{V}(\qa,\qb,\eta;\kjet,y_J) = \mathcal{K}_r^{(B)}\left(\qa,-\qb\right) \, \del{\qa+\qb-\kjet}\,\delta(\eta-y_J). 
\label{eq:jetvertexloy}
\end{equation}
Eqs. \eqref{eq:masterformula1} and \eqref{eq:jetvertexloy} will be the starting point for the NLO jet production in the symmetric configurations. 

Let us now switch to the asymmetric case. In general we can write $q_a$ and $q_b$ as 
\begin{align}
  q_a=&\alpha_a p_A+\beta_a p_B + q_{a \perp} & q_b=&\alpha_b p_A+\beta_b p_B + q_{b \perp} .
\end{align}
The strong ordering in the rapidity of emissions translates into the conditions $\alpha_a\gg\alpha_b$ and $\beta_b\gg\beta_a$. This, together with momentum 
conservation $q_a+q_b=k_J$, leads us to $\alpha_J=\alpha_a+\alpha_b\approx
\alpha_a$, $\beta_J=\beta_a+\beta_b\approx\beta_b$ and
\begin{align}
  s_{AJ} =& \beta_J s, & s_{BJ}=&\alpha_J s.
\label{eq:sajsbjfirst}
\end{align}
While the longitudinal momentum of $q_a (q_b)$ is a linear combination of 
$p_A$ and $p_B$ we see that only its component along $p_A (p_B)$ matters. 

If the colliding external particles provide no perturbative scale as it is 
the case in hadron--hadron collisions, then the jet is the only hard scale 
in the process and we have to deal with an asymmetric situation. Thus the 
scales $s_0$ and $s_0'$ should be chosen as $\kjet^2$ alone. At LO accuracy $s_0$ is arbitrary and we are indeed free to make this choice.
Then the arguments of the gluon Green's functions can be written as
\begin{align}
  \label{eq:sajs0lo}
  \frac{s_{AJ}}{s_0}=&\frac{1}{\alpha_a}, &
  \frac{s_{BJ}}{s_0}=&\frac{1}{\beta_b}. 
\end{align}
The description in terms of these longitudinal components is particularly 
useful if one is interested in jet production in a hadronic environment. 
Here one can introduce the concept of {\it unintegrated gluon 
density} in the hadron. This represents the probability of resolving a gluon 
carrying a longitudinal momentum fraction $x$ from the incoming hadron, and 
with a certain transverse momentum $k_T$. 
With the help of Eq.~\eqref{eq:sajs0lo} a LO unintegrated gluon distribution $g$ can be defined from Eq.~\eqref{eq:masterformula0} as
\begin{equation}
  \label{eq:updflo}
  g(x,\kpure) = \int\frac{d^2 {\bf q}}{2\pi {\bf q}^2}\,\Phi_{P}({\bf q})\,\int_{\delta-i\infty}^{\delta+i\infty}\frac{d\omega}{2\pi i}\, x^{-\omega} f_\omega({\bf q},\kpure).
\end{equation}

Then we can rewrite Eq.~\eqref{eq:masterformula0} as
\begin{multline}
\label{eq:masterformula2}
\frac{d\sigma}{d^{2}\kjet dy_J} = \int d^2 \qa\int dx_1 \int d^2 \qb\int  dx_2\;\\
\times g(x_1,\qa)g(x_2,\qb)\mathcal{V}(\qa,x_1,\qb,x_2;\kjet,y_J),
\end{multline}
with the LO jet vertex for the asymmetric situation being
\begin{multline}
\label{eq:jetvertexlo}
\mathcal{V}(\qa,x_1,\qb,x_2;\kjet,y_J)=\mathcal{K}_r^{(B)}\left(\qa,-\qb\right)\\
\times \del{\qa+\qb-\kjet}\,\delta\left(x_1-\sqrt{\frac{\kjet^2}{s}}e^{y_J}\right)\delta\left(x_2-\sqrt{\frac{\kjet^2}{s}}e^{-y_J}\right).
\end{multline}
Having presented our framework for the LO case, in both $\gamma^*\gamma^*$ and 
hadron--hadron collisions, we now proceed to explain 
in detail what corrections are needed to define our cross sections at NLO. 
Special attention should be put into the treatment of those scales with do not 
enter the LO discussion but are crucial at higher orders. 

\section{Inclusive jet production at NLO}

A similar approach to that shown in Section 4 remains valid when jet 
production is considered at NLO. The crucial step in this direction is 
to modify  the LO jet vertex of Eq.~\eqref{eq:jetvertexloy} and 
Eq.~\eqref{eq:jetvertexlo} to include new configurations present 
at NLO. We show how this is done in the following first subsection. 
In the second subsection we implement this vertex in the symmetric 
$\gamma^* \gamma^*$ case, and we repeat the steps from Eq.~\eqref{eq:total} to 
Eq.~\eqref{eq:sajsbjfirst}, carefully describing the choice of energy scale at 
each of the subchannels. In the third subsection hadron--hadron scattering is 
taken into consideration, and we extend the concept of unintegrated gluon 
density of Eq.~\eqref{eq:updflo} to NLO accuracy. Most importantly, it is 
shown that a correct choice of intermediate energy scales in this case 
implies a modification of the impact factors, the jet vertex, and the evolution kernel.

\subsection{The NLO jet vertex}

For those 
parts of the NLO kernel responsible for one gluon production we proceed in exactly the 
same way as at LO. The treatment of those terms related to two particle production is more complicated since  
for them it is necessary to introduce a jet algorithm. In general terms, if the two 
emissions generated by the kernel are nearby in phase space they will be considered as 
one single jet, otherwise one of them will be identified as the jet whereas the other 
will be absorbed as an untagged inclusive contribution. Hadronization effects in the 
final state are neglected and we simply define a cone of radius $R_0$ in the 
rapidity--azimuthal angle space such that two particles form a single jet if 
$R_{12} \equiv \sqrt{(\phi_1-\phi_2)^2+(y_1-y_2)^2} < R_0$. As long as only 
two emissions are involved this is equivalent to the $k_T$--clustering algorithm. 

To introduce the jet definition in the $2 \rightarrow 2$ components of the 
kernel it is convenient to start by considering the gluon and quark matrix 
elements together:
\begin{eqnarray}
\label{eq:k2q2g}
\left(\mathcal{K}_{\rm QMRK} + \mathcal{K}_{Q\bar{Q}} \right)(\qa,-\qb)  &=& 
\int\dktwo \int dy_2 \nonumber\\
&&\hspace{-3.8cm} \times \Big(\agsquare{\qa}{\qb}{\kone}{\ktwo}\theta(s_\Lambda-s_{12}) +\aqsquare{\qa}{\qb}{\kone}{\ktwo}\Big),
\end{eqnarray}
with ${\cal A}_{2P}$ being the two particle production amplitudes of which
 only the gluonic one also contributes to MRK. This is why a step function is 
needed to separate it from MRK. Momentum conservation implies that 
$\kone = \qa + \qb - \ktwo$.

The expression~\eqref{eq:k2q2g} is not complete  as it stands since we should 
also include the MRK contribution as it was previously done in 
Eq.~\eqref{eq:krrgg}: 
\begin{align}
&\left( {\cal K}_{GG} + {\cal K}_{Q\bar Q} \right) (\qa,-\qb) \equiv 
\int\dktwo\int dy_2\,\bsquare{\qa}{\qb}{\kone}{\ktwo} \nonumber\\
 =& \int\dktwo\int dy_2\,
\Bigg\{\agsquare{\qa}{\qb}{\kone}{\ktwo}\theta(s_\Lambda-s_{12})\nonumber\\
&-{\cal K}^{(B)}(\qa,\qa-\kone) \, {\cal K}^{(B)}(\qa-\kone,-\qb)\;\frac{1}{2}\,\theta\left(\ln\frac{s_{\Lambda}}{\ktwo^2}-y_2\right)\theta\left(y_2-\ln\frac{\kone^2}{s_{\Lambda}}\right)\nonumber\\
& +\aqsquare{\qa}{\qb}{\kone}{\ktwo}\Bigg\}. 
\label{eq:defbsquare}
\end{align}
We are now ready to introduce the jet definition for the double emissions. The 
NLO versions of Eq.~\eqref{eq:jetvertexloy} and Eq.~\eqref{eq:jetvertexlo} then read, respectively,
\begin{align}
\mathcal{V}(\qa,\qb,\eta;\kjet,y_J)= & \left(\mathcal{K}_r^{(B)} +\mathcal{K}_r^{(v)}\right)(\qa,-\qb) \Big|_{(a)}^{[y]}\non
  &\hspace{-2cm}+ \int\dktwo\;dy_2\bsquare{\qa}{\qb}{\kjet-\ktwo}{\ktwo}\theta(R_0-R_{12})\Big|_{(b)}^{[y]}\non
  &\hspace{-2cm}+ 2\int\dktwo\;dy_2\bsquare{\qa}{\qb}{\kjet}{\ktwo}\theta(R_{J2}-R_0)\Big|_{(c)}^{[y]},\label{eq:jetvertexnloy}\\
\mathcal{V}(\qa,x_1,\qb,x_2;\kjet,y_J)= & \left(\mathcal{K}_r^{(B)} +\mathcal{K}_r^{(v)}\right)(\qa,-\qb) \Big|_{(a)}^{[x]}\non
  &\hspace{-2cm}+ \int\dktwo\;dy_2\bsquare{\qa}{\qb}{\kjet-\ktwo}{\ktwo}\theta(R_0-R_{12})\Big|_{(b)}^{[x]}\non
  &\hspace{-2cm}+ 2\int\dktwo\;dy_2\bsquare{\qa}{\qb}{\kjet}{\ktwo}\theta(R_{J2}-R_0)\Big|_{(c)}^{[x]}.\label{eq:jetvertexnlox}
\end{align}
In these two expressions we have introduced the notation 
\begin{align}
  &\Big|_{(a,b)}^{[y]}&&\hspace{-1.3cm}=\del{\qa+\qb-\kjet} \delta (\eta - y^{(a,b)}),  \\
  &\Big|_{(c)}^{[y]}&&\hspace{-1.3cm}=\del{\qa+\qb-\kjet-\ktwo} \delta \left(\eta-y^{(c)}\right), \\
   &\Big|_{(a,b)}^{[x]}&&\hspace{-1.3cm}=\del{\qa+\qb-\kjet} \delta \left(x_1 - x_1^{(a,b)}\right)\delta \left(x_2 - x_2^{(a,b)}\right),  \\
  &\Big|_{(c)}^{[x]}&&\hspace{-1.3cm}=\del{\qa+\qb-\kjet-\ktwo} \delta \left(x_1 - x_1^{(c)}\right)   \delta \left(x_2 - x_2^{(c)}\right) .
\end{align}
The various jet configurations demand different $y$ and $x$ configurations. These are related to the properties of the produced jet in different ways 
depending on the origin of the jet: if only one gluon was produced in MRK this 
corresponds to the configuration (a) in the table below, if two particles in 
QMRK form a jet then we have the case (b), and finally case (c) if the jet is 
produced out of one of the partons in QMRK. The factor of 2 in the last term 
of Eq.~\eqref{eq:jetvertexnloy} and Eq.~\eqref{eq:jetvertexnlox} accounts for the possibility that either emitted 
particle can form the jet. Just by kinematics we get the explicit expressions for the different $x$ configurations listed in the following table:
\begin{center}
\begin{tabular}[h!]{c|c|cc}
  JET  & $y$ configurations & \multicolumn{2}{c}{$x$ configurations}
\\\hline
a) \raisebox{-1ex}{\includegraphics[height=.7cm]{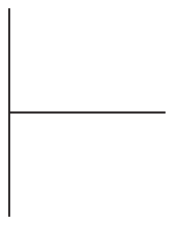}}  & 
   $y^{(a)}=y_J$ & $x_1^{(a)}=\frac{|\kjet|}{\sqrt{s}}e^{y_J}$ & 
   $x_2^{(a)}=\frac{|\kjet|}{\sqrt{s}}e^{-y_J}$ \\
b) \raisebox{-1.5ex}{\includegraphics[height=.7cm]{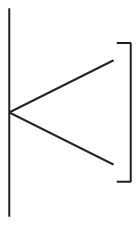}} & 
   $y^{(b)}=y_J$ & $x_1^{(b)}=\frac{\sqrt{\Sigma}}{\sqrt{s}}e^{y_J}$ & 
   $x_2^{(b)}=\frac{\sqrt{\Sigma}}{\sqrt{s}}e^{-y_J}$\\
c) \raisebox{-2ex}{\includegraphics[height=.7cm]{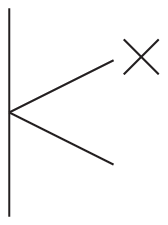}} & 
   $y^{(c)}=\frac{1}{2}\ln\frac{x_1^{(c)}}{x_2^{(c)}}$ & $ x_1^{(c)}=\frac{|\kjet|}{\sqrt{s}}e^{y_J}+\frac{|\ktwo|}{\sqrt{s}}e^{y_2}$ &   {\small $ x_2^{(c)}=\frac{|\kjet|}{\sqrt{s}}e^{-y_J}+\frac{|\ktwo|}{\sqrt{s}}e^{-y_2}$}
\end{tabular}
\end{center}
The variable $\Sigma$ is defined below in Eq.~\eqref{eq:defsigma}. Due to 
the analogue treatment of the emission vertex either expressed in terms of 
rapidities or longitudinal momentum fractions in the remaining of this 
section we will imply the same analysis for both. In particular, we will not 
explicitly mention these arguments when we come to 
Eqs.~(\ref{eq:jetvertexnlo2}, \ref{eq:freeofsing}). 

The introduction of the jet definition divides the phase space into 
different sectors. It is now needed to show that the final result is 
indeed free of any infrared divergences. In the following we proceed to 
independently calculate several contributions to the kernel to be able, in 
this way, to study its singularity structure.

The NLO virtual correction to the one--gluon emission kernel, 
${\cal K}^{(v)}$, was originally calculated in 
Ref.~\cite{Fadin:1993wh,Fadin:1994fj,Fadin:1996yv}. Its expression reads
\begin{eqnarray}
  \mathcal{K}_r^{(v)} \left(\qa,-\qb\right)&=& 
\frac{\bar{g}_\mu^4 \mu^{-2\epsilon}}{\pi^{1+\epsilon}\Gamma(1-\epsilon)}\frac{4}{\delt^2}\Bigg\{ 2\left(\frac{\delt^2}{\mu^2}\right)^\epsilon \left(-\frac{1}{\epsilon^2}+\frac{\pi^2}{2}-2 \, \epsilon \, \zeta(3)\right) \nonumber\\
&&\hspace{-2cm}+\frac{\beta_0}{N_c}\frac{1}{\epsilon}+\,\frac{3\delt^2}{\qa^2-\qb^2}\ln\left(\frac{\qa^2}{\qb^2}\right)
-\ln^2\left(\frac{\qa^2}{\qb^2}\right) \nonumber\\
&&\hspace{-2cm}+\,\left(1-\frac{n_f}{N_c}\right)\bigg[
\frac{\delt^2}{\qa^2-\qb^2}\left(1-\frac{\delt^2(\qa^2+\qb^2-4 \, \qa\qb)}{3(\qa^2-\qb^2)^2}\right)\ln\left(\frac{\qa^2}{\qb^2}\right)\nonumber\\
&&\hspace{-2cm}-\,\frac{\delt^2}{6\qa^2\qb^2}(\qa-\qb)^2+\frac{\delt^4 \, (\qa^2+\qb^2)}{6\qa^2\qb^2(\qa^2-\qb^2)^2}(\qa^2+\qb^2-4 \, \qa\qb)\bigg]\Bigg\},
\label{eq:kernelv}
\end{eqnarray}
with $\beta_0 = (11N_c-2n_f)/3$, $\zeta(n)= \sum_{k=1}^\infty k^{-n}$ and 
${\bf \Delta} = {\bf q}_a + {\bf q}_b$. ${\bar g}_\mu$ can be expressed in terms of 
the  renormalized coupling constant $g_\mu$ in the $\overline{\rm MS}$ renormalization scheme by the relation $\bar{g}_\mu^2 = g_\mu^2 \, N_c \, \Gamma(1-\epsilon) \, (4\pi)^{-2-\epsilon}$. Note that the expression for the virtual contribution given in \cite{Ostrovsky:1999kj} lacks the log squared.

Those pieces related to two--gluon production in QMRK can be rewritten 
in terms of their corresponding matrix elements as
\begin{eqnarray}
  \mathcal{K}_{\rm QMRK}(\qa,-\qb) &=& \int\dktwo\int dy_2 \agsquare{\qa}{\qb}{\kone}{\ktwo} \theta(s_\Lambda-s_{12})\nonumber\\
&&\hspace{-2cm}   = \frac{g_\mu^2 \mu^{-2\epsilon}N_c^2}{\pi(2\pi)^{D+1}\qa^2\qb^2}\int\dktwoeps\int dy_2\; A_{\text{gluons}}\,\theta(s_\Lambda-s_{12}),
\label{eq:kernelagluons}
\end{eqnarray}
and those related to quark--antiquark production are
\begin{eqnarray}
\mathcal{K}_{Q\bar{Q}}(\qa,-\qb) &=& \int\dktwo\int dy_2 \aqsquare{\qa}{\qb}{\kone}{\ktwo} \nonumber\\
   &=& \frac{g_\mu^2 \mu^{-2\epsilon}N_c^2}{\pi(2\pi)^{D+1}\qa^2\qb^2}\int\dktwoeps\int dy_2\; A_{\text{quarks}}\label{eq:kernelaquarks}.
\end{eqnarray}
We have calculated the corresponding amplitudes, using the Mandelstam 
invariants $\shat$, $\that$, and $\uhat$, and our results are 
\begin{eqnarray}
  A_{\text{gluons}}&=& \qa^2\qb^2 \Bigg\{-\frac{1}{\that\uhat}+\frac{1}{4\that\uhat}\frac{\qa^2\qb^2}{\kone^2\ktwo^2}-\frac{1}{4}\left(\frac{1-x}{x}\frac{1}{\ktwo^2\that}+\frac{x}{1-x}\frac{1}{\kone^2\uhat}\right)+\frac{1}{4\kone^2\ktwo^2}\nonumber\\
&& \hspace{0.1cm}+\frac{1}{\Sigma}\Bigg[-\frac{1}{\shat}\left(2+\left(\frac{1}{\that}-\frac{1}{\uhat}\right)\left(\frac{1-x}{x}\kone^2-\frac{x}{1-x}\ktwo^2\right)\right)+\frac{1}{4}\left(\frac{\Sigma}{\shat}+1\right)\nonumber\\
&& \hspace{1cm}\times\left(\frac{1-x}{x}\frac{1}{\ktwo^2}+\frac{x}{1-x}\frac{1}{\kone^2}\right) -\frac{\qb^2}{4\shat}\left(\frac{1}{(1-x)\that}+\frac{1}{x\uhat}\right)\nonumber\\
&& \hspace{1cm}-\frac{\qa^2}{4\shat}\left(\left[1+\frac{x}{1-x}\frac{\ktwo^2}{\kone^2}\right]\frac{1}{\that}+\left[1+\frac{1-x}{x}\frac{\kone^2}{\ktwo^2}\right]\frac{1}{\uhat}\right)\Bigg]\Bigg\}\nonumber\\
&+&\frac{D-2}{4}\Bigg\{\left(\frac{(\kone-\qa)^2(\ktwo-\qa)^2-\kone^2\ktwo^2}{\that\uhat}\right)^2\nonumber\\
&& \hspace{-.5cm}-\frac{1}{4}\left(\frac{(\ktwo-\qa)^2-\frac{x}{1-x}\ktwo^2}{\uhat}+\frac{E}{\shat}\right)
\left(\frac{(\kone-\qa)^2-\frac{1-x}{x}\kone^2}{\that}-\frac{E}{\shat}\right)\Bigg\},
\label{eq:Agluons}
\end{eqnarray}
\begin{eqnarray}
A_{\text{quarks}} &=& \frac{n_f}{4N_c}\Bigg\{\frac{\qa^2\qb^2}{\shat\Sigma}\left(2+\left(\frac{1}{\that}-\frac{1}{\uhat}\right)\left(\frac{1-x}{x}\kone^2-\frac{x}{1-x}\ktwo^2\right)\right)\nonumber\\
&&-\left(\frac{(\kone-\qa)^2(\ktwo-\qa)^2-\kone^2\ktwo^2}{\that\uhat}\right)^2\nonumber\\
&&+\frac{1}{2}\left(\frac{(\ktwo-\qa)^2-\frac{x}{1-x}\ktwo^2}{\uhat}+\frac{E}{\shat}\right)
\left(\frac{(\kone-\qa)^2-\frac{1-x}{x}\kone^2}{\that}-\frac{E}{\shat}\right)\Bigg\}\nonumber\\
&&+\frac{n_f}{4N_c^3}\Bigg\{\left(\frac{(\kone-\qa)^2(\ktwo-\qa)^2-\kone^2\ktwo^2}{\that\uhat}\right)^2-\frac{\qa^2\qb^2}{\that\uhat}\Bigg\} .
\end{eqnarray}
These expressions are in agreement with the corresponding ones obtained in 
Ref.~\cite{Ostrovsky:1999kj}. The following notation has been used:
\begin{eqnarray}
  x &=& \frac{|\kone|}{|\kone|+|\ktwo|e^{\Delta y}}, \\
 \lambd &=& (1-x)\kone-x\ktwo, \\
\Sigma &=& \shat+\delt^2 \, = \, \frac{\lambd^2}{x(1-x)} + {\bf \Delta}^2 , \label{eq:defsigma}\\
 E &=& 2(2x-1)\qa^2+4\lambd\qa +\frac{1-2x}{x(1-x)}\lambd^2 \nonumber\\
&-& 2 x (1-x)\left((2x-1)\delt^2+2\lambd\delt\right)\frac{\qa^2}{x(1-x)\delt^2+\lambd^2}
\label{eq:e}.
\end{eqnarray}
We now study those terms which contribute to generate soft and collinear divergences after integration over the two--particle phase space. They should be able to cancel the $\epsilon$ poles of the virtual contributions in Eq.~\eqref{eq:kernelv}, {\it i.e.}
\begin{eqnarray}
  \mathcal{K}^{(v)}_{\rm singular} \left(\qa,\qb\right)&=& 
\frac{\bar{g}_\mu^4 \mu^{-2\epsilon}}{\pi^{1+\epsilon}\Gamma(1-\epsilon)}\frac{4}{\delt^2}\Bigg\{\left(\frac{\delt^2}{\mu^2}\right)^\epsilon \left(-\frac{2}{\epsilon^2}\right) +\frac{\beta_0}{N_c}\frac{1}{\epsilon}\Bigg\}.
\label{eq:kernelvsingular}
\end{eqnarray}
Here we identify those pieces responsible for the generation of these poles.

One of the divergent regions is defined by the two emissions with momenta 
$k_1=\alpha_1 p_A+\beta_1 p_B +k_{1\perp}$ and $k_2=\alpha_2 p_A+\beta_2 p_B +k_{2\perp}$ becoming collinear. This means that, for a real parameter $\lambda$, $k_1 \simeq \lambda \, k_2$, {\it i.e.} $k_{1\perp} \simeq \lambda \, k_{2\perp}$, $\alpha_1 \simeq \lambda \, \alpha_2$ and thus $\alpha_2 k_{1\perp} - \alpha_1 k_{2\perp} \simeq 0$. Since $x=\frac{\alpha_1}{\alpha_1+\alpha_2}$ this is equivalent to the condition $\lambd \simeq 0$. In the collinear region $\shat=\frac{\lambd^2}{x(1-x)}$ tends to zero and the dominant contributions which are 
purely collinear are 
\begin{align}
  A_{\text{gluons}}^{\text{singular}}\Big|_{\rm collinear} =& 
- \frac{\qa^2\qb^2}{\Sigma}\frac{2}{\shat} +\frac{D-2}{16}\frac{E^2}{\shat^2} \equiv A_{(1)}+A_{(2)}\label{eq:agluonscollinear},\\
  A_{\text{quarks}}^{\text{singular}}\Big|_{\rm collinear} =& \frac{n_f}{2N_c}\frac{\qa^2\qb^2}{\shat\Sigma}-\frac{n_f}{8N_c}\frac{E^2}{\shat^2} \label{eq:aquarkscoll}.
\end{align}
The quark--antiquark production does not generate divergences when 
${\bf k}_1$ or ${\bf k}_2$ become soft, therefore we have that the only 
purely soft divergence is
\begin{equation}
  A_{\text{gluons}}^{\text{singular}}\Big|_{\rm soft} = \qa^2\qb^2\Bigg(\frac{1}{4\that\uhat}\frac{\qa^2\qb^2}{\kone^2\ktwo^2}+\frac{1}{4\kone^2\ktwo^2}\Bigg) \equiv A_{(3)}+A_{(4)}
~\rightarrow~ 2A_{(4)},
\label{eq:agluonssoft}
\end{equation}
where we have used the property that, in the soft limit, the $\that\uhat$ 
product tends to $\qa^2\qb^2$. We will see that these terms will be 
responsible for simple poles in $\epsilon$. The double poles will be generated 
by the regions with simultaneous soft and collinear divergences. They are 
only present in the gluon--gluon production case and can be written as
\begin{eqnarray}
  A_{\text{gluons}}^{\text{singular}}\Big|_{\rm soft \& collinear} &=& 
\frac{\qa^2\qb^2}{4\shat}\left[\frac{1-x}{x}\frac{1}{\ktwo^2}
+\frac{x}{1-x}\frac{1}{\kone^2}\right] \nonumber\\
&&\hspace{-4.1cm}-\frac{\qa^2\qb^2}{4\shat\Sigma}
\Bigg[\qb^2\left(\frac{1}{(1-x)\that}+\frac{1}{x\uhat}\right)
+\qa^2\left(\left[1+\frac{x}{1-x}\frac{\ktwo^2}{\kone^2}\right]\frac{1}{\that}
+\left[1+\frac{1-x}{x}\frac{\kone^2}{\ktwo^2}\right]
\frac{1}{\uhat}\right)\Bigg].\nonumber\\
&=&A_{(5)}+A_{(6)}.\label{eq:agluonssoftcoll}
\end{eqnarray}
Focusing on the divergent structure it turns out that in the soft and 
collinear region the first line of Eq.~\eqref{eq:agluonssoftcoll}, $A_{(5)}$,  
has exactly the same limit as the second line, $A_{(6)}$. This is very 
convenient since we can then simply write
\begin{eqnarray}
  A_{\text{gluons}}^{\text{singular}}\Big|_{\rm soft \& collinear} 
&\rightarrow& 
\frac{\qa^2\qb^2}{2\shat}\left(\frac{1-x}{x}\frac{1}{\ktwo^2}+\frac{x}{1-x}\frac{1}{\kone^2}\right)=2A_{(5)}.\label{eq:agluonssoftcoll2}
\end{eqnarray}
The MRK contribution of Eq.~\eqref{eq:defbsquare} has the form 
$A_{\rm MRK}= - 4 A_{(4)}$ and when added to all the other singular terms we 
get the expression
\begin{eqnarray}
\int\dktwo\int dy_2\,\bssquare{\qa}{\qb}{\ktwo}{\kone} &\equiv& \nonumber\\
&&\hspace{-7cm}
\frac{g_\mu^2 \mu^{-2\epsilon}N_c^2}{\pi(2\pi)^{D+1}\qa^2\qb^2}
\int\dktwoeps \int dy_2 \, \Big\{A^{\rm singular}_{\text{gluons}} \, 
\theta(s_\Lambda-s_{12})+ A^{\rm singular}_{\text{quarks}}\Big\}
\label{eq:defbssquare},
\end{eqnarray}
with
\begin{eqnarray}
A^{\rm singular}_{\text{gluons}} \, 
\theta(s_\Lambda-s_{12})+ A^{\rm singular}_{\text{quarks}} &=& \nonumber\\
&&\hspace{-6cm}\left\{\underbrace{- \frac{\qa^2\qb^2}{\Sigma}\frac{2}{\shat}}_{\rm Gluon|_{coll_1}} +\underbrace{\frac{D-2}{16}\frac{E^2}{\shat^2}}_{\rm Gluon|_{coll_2}}-\underbrace{\frac{\qa^2\qb^2}{2\kone^2\ktwo^2}}_{\rm Gluon|_{soft}}+\underbrace{\frac{\qa^2\qb^2}{2\shat}\left(\frac{1-x}{x}\frac{1}{\ktwo^2}+\frac{x}{1-x}\frac{1}{\kone^2}
\right)}_{\rm Gluon|_{\rm soft \& coll}}\right\} \theta(s_\Lambda-s_{12}) \nonumber\\
&&\hspace{-3cm}+ \underbrace{\frac{n_f}{2N_c}\frac{\qa^2\qb^2}{\shat\Sigma}}_{\rm Quark|_{coll_1}}-\underbrace{\frac{n_f}{8N_c^3}\frac{E^2}{\shat^2}}_{\rm Quark|_{coll_2}}.\hspace{-3cm}.
\label{eq:differentterms}
\end{eqnarray}
We have labeled the different terms to study how each of them produces the 
$\epsilon$ poles. We will do this in Section 6.

With the singularity structure well identified we now come back to 
Eqs.~(\ref{eq:jetvertexnloy}, \ref{eq:jetvertexnlox}) and show how they are free of any divergences.
Only if the divergent terms belong to the same 
configuration this cancellation can be shown analytically. With this in mind we 
add the singular parts of the two particle production of 
Eq.~\eqref{eq:defbssquare} in the configuration $(a)$ 
multiplied by $0=1-\theta(R_0-R_{12})-\theta(R_{12}-R_0)$:
\begin{eqnarray}
\mathcal{V} &=& \bigg[\left(\mathcal{K}_r^{(B)}
+\mathcal{K}_r^{(v)}\right)(\qa,-\qb) \nonumber\\
&&\hspace{3cm}+\int\dktwo\;dy_2\bssquare{\qa}{\qb}{\kjet-\ktwo}{\ktwo}\bigg] \Big|_{(a)}\nonumber\\
&+& \int\dktwo\;dy_2\bigg[\bsquare{\qa}{\qb}{\kjet-\ktwo}{\ktwo}\Big|_{(b)}\nonumber\\
&&\hspace{3cm}-\bssquare{\qa}{\qb}{\kjet-\ktwo}{\ktwo}\Big|_{(a)}\bigg]\theta(R_0-R_{12})\nonumber\\
&+&\bigg[2\int\dktwo\;dy_2\bsquare{\qa}{\qb}{\kjet}{\ktwo}\theta(R_{J2}-R_0)\Big|_{(c)}\nonumber\\
&&\hspace{0.4cm}-\int\dktwo\;dy_2\bssquare{\qa}{\qb}{\kjet-\ktwo}{\ktwo}\theta(R_{12}-R_0)\Big|_{(a)}\bigg].
\label{eq:jetvertexnlo2}
\end{eqnarray}

The cancellation of divergences within the first two lines is now the same 
as in the calculation of the full NLO kernel. In Section 6 we will show how the first two lines of 
Eq.~\eqref{eq:jetvertexnlo2} are free of any singularities in the form of 
$\epsilon$ poles. In doing so we will go into the details of the r\^ole of $s_\Lambda$.
The third and fourth lines are also explicitly free 
of divergences since these have been subtracted out. The sixth line has a 
${\bf k}_1 \leftrightarrow {\bf k}_2$ symmetry which allows us to write
\begin{eqnarray}
\mathcal{V} &=& \bigg[\left(\mathcal{K}_r^{(B)}+\mathcal{K}_r^{(v)}\right)(\qa,-\qb)\nonumber\\
&&\hspace{3cm} +\int\dktwo\;dy_2\bssquare{\qa}{\qb}{\kjet-\ktwo}{\ktwo}\bigg] \Big|_{(a)}\nonumber\\
&+& \int\dktwo\;dy_2\bigg[\bsquare{\qa}{\qb}{\kjet-\ktwo}{\ktwo}\Big|_{(b)}\nonumber\\
&&\hspace{3cm}-\bssquare{\qa}{\qb}{\kjet-\ktwo}{\ktwo}\Big|_{(a)}\bigg]\theta(R_0-R_{12})\nonumber\\
&+&2\int\dktwo\;dy_2\bigg[\bsquare{\qa}{\qb}{\kjet}{\ktwo}\theta(R_{J2}-R_0)\Big|_{(c)}\nonumber\\
&&\hspace{0.4cm}-\bssquare{\qa}{\qb}{\kjet-\ktwo}{\ktwo}\theta(R_{12}-R_0)\theta(|\kone|-|\ktwo|)\Big|_{(a)}\bigg].
\label{eq:freeofsing}
\end{eqnarray}
We can now see that the remaining possible divergent regions of the last line 
are regulated by the cone radius $R_0$. 

It is worth noting that, apart from an overall 
${\bar \alpha}_s^2 (\mu^2)$ factor, the NLO terms in the last four lines in 
Eq.~\eqref{eq:freeofsing} do not carry any renormalization scale dependence since 
they are finite when $\epsilon$ is set to zero. The situation is different for the 
first two lines since $\mathcal{V}$ contains a logarithm of $\mu^2$ in the form
\begin{equation}
  \label{eq:runningcoupling}
  \mathcal{V}= \mathcal{V}^{(B)}\left(1-\frac{\alpha_s(\mu^2)}{4\pi}\frac{\beta_0}{N_c}\ln\frac{\kjet^2}{\mu^2}\right) + \Delta\mathcal{V}.
\end{equation}
where $\Delta\mathcal{V}$ contains the third to sixth lines and the $\mu$--independent part of the first two lines of \eqref{eq:freeofsing}. It is then natural to absorb this term in a redefinition of the running of the 
coupling and replace $\alpha_s(\mu^2)$ by $\alpha_s(\kjet^2)$. For a explicit derivation 
of this term we refer the reader to Section 6.

Therefore we now have a finite expression for the jet vertex suitable for numerical 
integration. 
This numerical analysis will be performed elsewhere since here we are mainly 
concerned with the formal introduction of the jet definition 
and the correct separation of the different contributions to the kernel. 

What remains to be proven is the cancellation of divergences between 
Eq.~\eqref{eq:kernelvsingular} and Eq.~\eqref{eq:defbssquare}. This will be 
performed in Section 6. Before doing so, in the next two subsections, we indicate 
how to introduce our vertex in the definition of the differential cross 
section. Especial care must be taken in the treatment of the energy scale 
in the Reggeized gluon propagators since in the symmetric case it is directly  
related to the rapidity difference between subsequent emissions, as we will  
show in the next subsection, but in the asymmetric case of hadron--hadron 
collisions it depends on the longitudinal momentum fractions 
of the $t$--channel Reggeons. 

\subsection{Production of jets in $\gamma^*\gamma^*$ scattering}

We now have all the ingredients required to describe the inclusive single jet 
production in a symmetric process at NLO. To be definite, we consider $\gamma^*\gamma^*$ scattering with the 
virtualities of the two photons being large and of the same order. All we need is to 
take Eq.~\eqref{eq:masterformula0} for the differential cross section as a 
function of the transverse momentum and rapidity of the jet. The vertex 
${\cal V}$ to be used is that of Eq.~\eqref{eq:freeofsing} in the 
representation based on rapidity variables of Eq.~\eqref{eq:jetvertexnloy}. 
The rapidities of the emitted particles are the natural variables to 
characterize the partonic evolution and $s$--channel production since we 
assume that all transverse momenta are of the same order. 

Let us note that the rapidity difference between two emissions can be 
written as
\begin{eqnarray}
y_i - y_{i+1} &=& \ln{\frac{s_{i,i+1}}{\sqrt{{\bf k}_i^2 {\bf k}_{i+1}^2}}}
\end{eqnarray}
which supports the choice $s_{R;i,i+1}=\sqrt{{\bf k}_i^2{\bf k}_{i+1}^2}$ in 
Eq.~\eqref{eq:news17nlo}. This is also technically more convenient since it 
simplifies the final expression for the cross section in Eq.~\eqref{MRKfinal}.

\begin{center}
\begin{figure}[htbp]
  \centering
  \includegraphics[width=8cm]{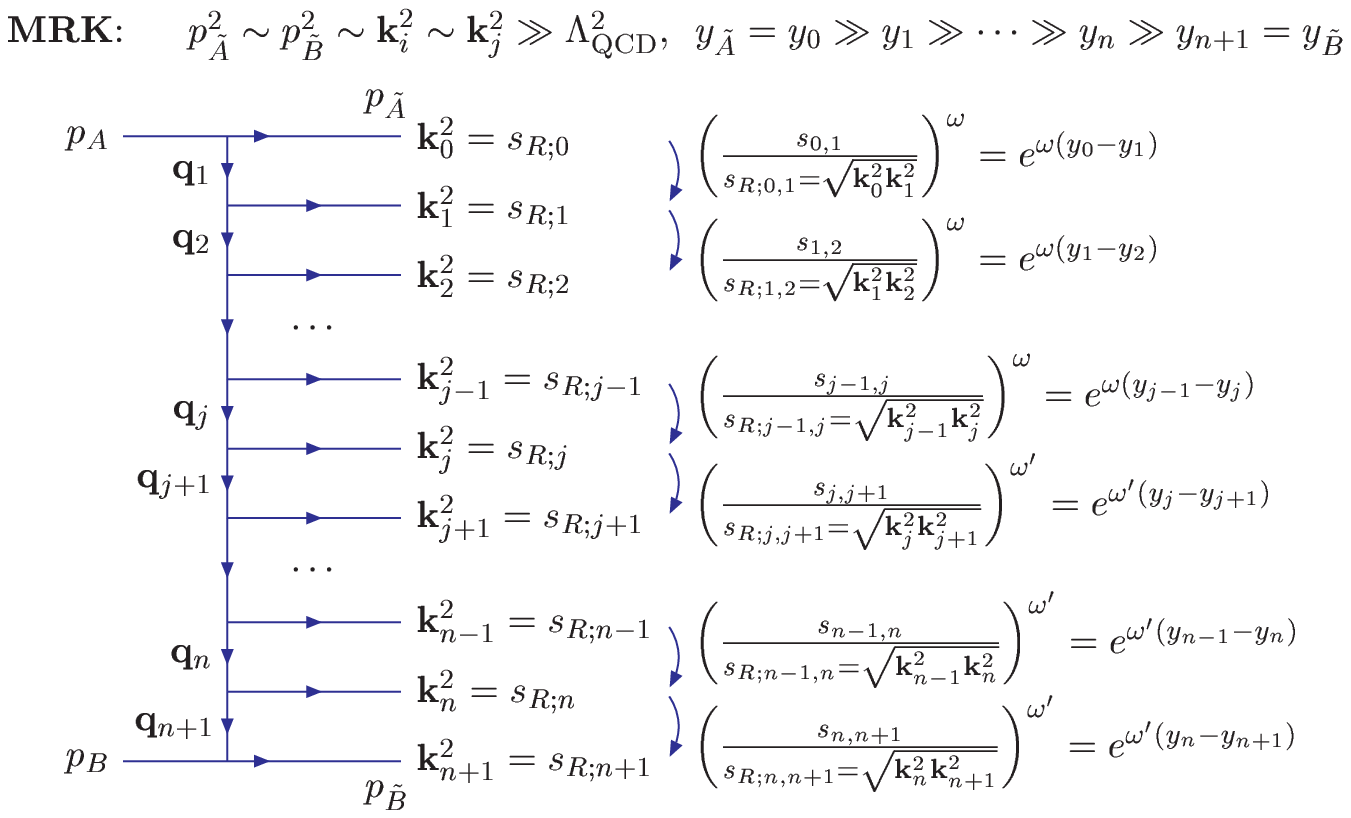}
  \caption{Momenta for $2 \rightarrow 2 + (n-1) + {\rm jet}$ amplitude 
in the symmetric configuration with MRK. The produced jet has rapidity 
$y_J=y_j$ and transverse momentum $\kjet=\kj$.}
  \label{SymmetricMRK}
\end{figure}
\end{center}

In Fig.~\ref{SymmetricMRK} we illustrate the different scales 
participating in the scattering and the variables of evolution. We write 
down the conditions for MRK: all transverse momenta are of similar size and 
much larger than the confining scale, the rapidities are strongly ordered in 
the evolution from one external particle to the other. At each stage of the 
evolution the propagation of the Reggeized gluons, which generates  
rapidity gaps, takes place between two real emissions. There are many 
configurations contributing to the differential cross section, each of them 
with a different weight. Eq.~\eqref{eq:masterformula0} represents the sum of
these production processes.

\subsection{The unintegrated gluon density and jet production in 
hadron--hadron collisions}

In this subsection we now turn to the case of hadron collisions where 
MRK has to be necessarily modified to include some evolution in the 
transverse momenta, since the momentum of the jet will be much 
larger than the typical transverse scale associated to the hadron.

In the LO case we have already explained that, in order to move from the symmetric 
case to the asymmetric one, it is needed to change the energy scale from 
Eq.~\eqref{sosoprime} to Eq.~\eqref{eq:sajs0lo}. This is equivalent to 
changing the description of the evolution in terms of rapidity 
differences between emissions to longitudinal momentum fractions of the
Reggeized gluons in the $t$--channel. 
Whereas in LO this change of scales has no consequences, in NLO accuracy it
leads to modifications, not only of the jet emission vertex but
also of the evolution kernels above and below the jet vertex. 
\begin{figure}
  \centering
  \includegraphics[width=7cm]{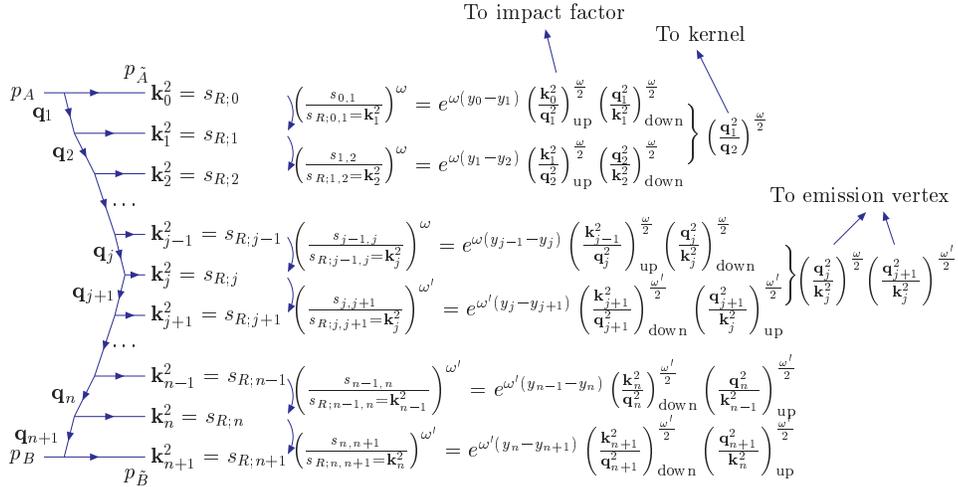}
  \caption{Momenta for $2 \rightarrow 2 + (n-1) + {\rm jet}$ amplitude 
in the asymmetric configuration with $k_t$--ordered MRK.}
  \label{AsymmetricMRK}
\end{figure}
These new definitions will allow the cross section still to be written in a factorizable way and the 
evolution of the gluon Green's function still to be described by an integral 
equation. 

To understand this in detail we start by writing the solution 
to the NLO BFKL equation iteratively, {\it i.e.}
\begin{eqnarray}
\int d^2 \ka f_\omega(\ka,\qa) &=& 
\frac{1}{\omega} \sum_{j=1}^\infty 
\left[\prod_{i=1}^{j-1}\int d^2 \qi\frac{1}{\omega}\mathcal{K}(\qi,\qip)\right],
\label{eq:ggfiterando}
\end{eqnarray}
where $\qone = \ka$ and $\qj = \qa $.
We now focus on one side of the 
evolution towards the hard scale since the other side is similar and use 
Fig.~\ref{AsymmetricMRK} as a graphical reference. Starting 
with the symmetric case the differential cross section for jet 
production contains the following evolution between particle $A$ and 
the jet:
\begin{eqnarray}
  \frac{d\sigma}{d^2 \kjet dy_J} &=& \int d^2 \qa \int d^2 \ka 
\frac{\Phi_A(\ka)}{2\pi\ka^2} \nonumber\\
&&\hspace{0.1cm}\times\int\frac{d\omega}{2\pi i} 
f_\omega(\ka,\qa)\left(\frac{s_{AJ}}{\sqrt{\ka^2\kjet^2}}\right)^\omega 
{\cal V}(\qa,\qb;\kjet,y_J)\dots
\end{eqnarray}

In the asymmetric situation where $\kjet^2\gg \ka^2$ the scale 
$\sqrt{\ka^2\kjet^2}$ should be replaced by $\kjet^2$. In order to do so 
we rewrite the term related to the choice of energy scale. To be consistent 
with Fig.~\ref{AsymmetricMRK} we take $\kj = \kjet$, 
${\bf k}_0 =  -\ka = -\qone$ and ${\bf q}_j = \qa$. To start 
with it is convenient to introduce a chain of scale changes in every kernel:
\begin{eqnarray}
 \left(\frac{s_{AJ}}{\sqrt{\ka^2\kjet^2}}\right)^\omega &=& 
\left[\prod_{i=1}^{j}\left(\frac{\ki^2}{\kim^2}\right)^{\frac{\omega}{2}}\right]  \left(\frac{s_{AJ}}{\kjet^2}\right)^\omega, 
\end{eqnarray}
which can alternatively be written in terms of the $t$--channel momenta as
\begin{eqnarray}
 \left(\frac{s_{AJ}}{\sqrt{\ka^2\kjet^2}}\right)^\omega &=& \left[\prod_{i=1}^{j-1}\left(\frac{\qip^2}{\qi^2}\right)^{\frac{\omega}{2}}\right]\left(\frac{\kjet^2}{\qa^2}\right)^{\frac{\omega}{2}}\left(\frac{s_{AJ}}{\kjet^2}\right)^\omega.
\end{eqnarray}
For completeness note that we are indeed changing the variable of evolution 
from a difference in rapidity:
\begin{eqnarray}
\frac{s_{AJ}}{\sqrt{\ka^2\kjet^2}} &=& e^{y_{\tilde A}-y_J} 
\end{eqnarray}
to the inverse of the longitudinal momentum fraction, {\it i.e.}
\begin{eqnarray}
\frac{s_{AJ}}{\kjet^2} &=& \frac{1}{\alpha_J}.
\label{jetinverselong}
\end{eqnarray}

This shift in scales translates into the following expression for the cross 
section:
\begin{eqnarray}
\frac{d\sigma}{d^2 \kjet dy_J} &=& 
\int\frac{d\omega}{2\pi i \, \omega}\sum_{j=1}^\infty 
\left[\prod_{i=1}^{j}\int d^2 \qi\right] \frac{\Phi_A(\qone)}{2\pi\qone^2}
\nonumber\\
&&\hspace{-2.4cm}\times\left[\prod_{i=1}^{j-1}\left(\frac{\qip^2}{\qi^2}\right)^{\frac{\omega}{2}}\frac{1}{\omega}\mathcal{K}(\qi,\qip)\right]\left(\frac{\kjet^2}{\qa^2}\right)^{\frac{\omega}{2}}
{\cal V}(\qa,\qb;\kjet,y_J)\left(\frac{s_{AJ}}{\kjet^2}\right)^\omega\ldots
\end{eqnarray}
As we mentioned above these changes can be absorbed at NLO in the kernels and 
impact factors, we just need to perturbatively expand the integrand. The 
impact factors get one single contribution, as can be seen in 
Fig.~\ref{AsymmetricMRK}, and they explicitly change as
\begin{eqnarray}
\label{newimpactfactor}
 \widetilde{\Phi}(\ka)&=& \Phi(\ka) -\frac{1}{2}{\ka^2}\int d^2 {\bf q} 
\frac{\Phi^{(B)}({\bf q})}{{\bf q}^2}\mathcal{K}^{(B)}({\bf q},\ka)
\ln\frac{{\bf q}^2}{\ka^2}.
\end{eqnarray}
The kernels in the evolution receive a double contribution from the different 
energy scale choices of both the incoming and outgoing Reggeons (see 
Fig.~\ref{AsymmetricMRK}). This amounts to the following correction:
\begin{eqnarray}
\label{newkernel}
  \widetilde{\mathcal{K}}(\qone,\qtwo) &=& \mathcal{K}(\qone,\qtwo)
-\frac{1}{2}\int d^2 {\bf q} \, \mathcal{K}^{(B)}(\qone,{\bf q}) 
\, \mathcal{K}^{(B)}({\bf q},\qtwo)\ln\frac{{\bf q}^2}{\qtwo^2}.
\end{eqnarray}
There is a different type of term in the case of the emission vertex 
where the jet is defined. This correction has also two contributions 
originated at the two different evolution chains from the hadrons $A$ and 
$B$. Its expression is
\begin{eqnarray}
\label{newemissionvertex}
  \widetilde{\cal V}(\qa,\qb) &=& {\cal V}(\qa,\qb)
-\frac{1}{2}\int d^2 {\bf q} \,  
\mathcal{K}^{(B)}(\qa,{\bf q}) {\cal V}^{(B)}({\bf q},\qb)
\ln\frac{{\bf q}^2}{({\bf q}-\qb)^2}\nonumber\\
&&\hspace{1cm}-\frac{1}{2}\int d^2 {\bf q} \, {\cal V}^{(B)}(\qa,{\bf q}) \,
\mathcal{K}^{(B)}({\bf q},\qb)\ln\frac{{\bf q}^2}{(\qa-{\bf q})^2}.
\end{eqnarray}

These are all the modifications we need to be able to write our differential 
cross section for the asymmetric case. The final expression is
\begin{eqnarray}
  \frac{d\sigma}{d^2 \kjet dy_J} &=& \int d^2 \qa\int d^2 \ka 
\frac{\widetilde{\Phi}_A(\ka)}{2\pi\ka^2}\nonumber\\
&&\hspace{1cm}\times \int\frac{d\omega}{2\pi i}
\tilde{f}_\omega(\ka,\qa)\left(\frac{s_{AJ}}{\kjet^2}\right)^\omega 
\widetilde{\cal V}(\qa,\qb;\kjet,y_J)\ldots
\end{eqnarray}
As in the LO case, we can use Eq.~\eqref{jetinverselong} to define the 
NLO unintegrated gluon density as
\begin{equation}
  g(x,{\bf k}) = \int d^2 {\bf q}\frac{\widetilde{\Phi}_P({\bf q})}{2\pi{\bf q}^2}\int\frac{d\omega}{2\pi i}\tilde{f}_\omega({\bf k},{\bf q})\, x^{-\omega}.
\end{equation}
The gluon Green's function ${\tilde f}_\omega$ is the solution to a new 
BFKL equation with the modified kernel of Eq.~\eqref{newkernel} which includes the energy shift at 
NLO, {\it i.e.}
\begin{equation}
  \omega \tilde{f}_\omega(\ka,\qa) = \del{\ka-\qa} +
\int d^2{\bf q} \, \widetilde{\mathcal{K}}(\ka,{\bf q}) \,
\tilde{f}_\omega({\bf q},\qa).
\end{equation}
In this way the unintegrated gluon distribution follows the evolution equation 
\begin{equation}
  \frac{\partial g(x,\qa)}{\partial\ln 1/x} 
= \int d^2 {\bf q} \, \widetilde{\mathcal{K}} (\qa,{\bf q}) \, g(x,{\bf q}).
\end{equation}
Finally, taking into account the evolution from the other hadron, the 
differential cross section reads
 \begin{equation}
  \frac{d\sigma}{d^2 \kjet dy_J} = \int d^2 \qa \int d^2 \qb 
\, g(x_a,\qa) \, g(x_b,\qb) \, \widetilde{\cal V}(\qa,\qb;\kjet,y_J),
\label{ppfinal} 
\end{equation}
with the emission vertex taken from Eq.~\eqref{newemissionvertex}.

We would like to indicate that with the prescription derived in this 
subsection we managed to express the new kernels, emission vertex and 
impact factors as functions of their incoming momenta only. It is also 
worth mentioning that the proton impact factor contains 
non--perturbative physics which can only be modeled by, {\it e.g.}
\begin{eqnarray}
\Phi_P({\bf q}) &\sim& (1-x)^{p_1} x^{- p_2} 
\left(\frac{{\bf q}^2}{{\bf q}^2+Q_0^2}\right)^{p_3},
\label{eq:protonIF}
\end{eqnarray}
where $p_i$ are positive free parameters, with $Q_0^2$ representing a momentum
scale of the order of the 
confinement scale. The initial $x$ dependence in this expression would be of 
non--perturbative origin. 

Let us also point out that the prescription to modify the kernel as in 
Eq.~\eqref{newkernel} was originally suggested in the first paper of
Ref.~\cite{FLCC} in the context of deep inelastic scattering.  
This new kernel can be considered as the first term in 
an all orders perturbative expansion due to the change of scale. When all 
terms are included the kernel acquires improved convergence properties and 
matches collinear evolution. Details of this procedure can be found in 
Ref.~\cite{Ciafaloni:2003rd}, where the collinear resummation was done 
in Mellin space. In a future publication we intend to investigate  
how these corrections can be phrased in momentum space, and how they affect
the behaviour of the unintegrated gluon distribution. For this we will use the procedure developed in 
Ref.~\cite{Vera:2005jt} where the resummation to all orders corresponding 
to the energy shift was proven to be equivalent to a Bessel function of the 
first kind with argument depending on the strong coupling and a double
logarithm of the 
ratio of transverse scales.

\section{Cancellation of divergences and a closer look at the separation 
between MRK and QMRK}

During the calculation of a NLO BFKL cross section, both at a fully inclusive 
level and at a more exclusive one, there is a need to separate 
the contributions from MRK and QMRK. In order to do so we have followed 
Ref.~\cite{Fadin:1998sh} and introduced the parameter $s_\Lambda$ in 
Eq.~\eqref{eq:kqmrk} and Eq.~\eqref{eq:phiqmrk}. In principle, at NLO 
accuracy, our final results should not depend on this extra scale. 
In fact, as we have remarked earlier in our discussion of the total
cross section (after Eq.~\eqref{eq:krrgg}), 
we could have taken the limit $s_{\Lambda} \to \infty$: the logarithms 
of $s_{\Lambda}$ cancel, and the corrections to the finite pieces die 
away as ${\cal O}(s_\Lambda^{-1})$.
In the context of the inclusive cross section, however, we prefer to treat    
$s_{\Lambda}$ as a physical parameter: it separates MRK from QMRK and, hence, 
cannot be arbitrarily large. We will therefore retain the dependence upon 
$s_{\Lambda}$: in the remainder of this section we demonstrate that, 
in our inclusive cross section, all logarithmic terms cancel 
(analogous to Eq.~\eqref{eq:krrgg}), and we will 
then leave the study of the corrections of the order 
${\cal O}(s_\Lambda^{-1})$ for a numerical analysis. It will also be 
interesting to see how this dependence on $s_\Lambda$ could be 
related to the rapidity veto introduced in Ref.~\cite{RapidityVeto}. 

Let us consider the $s_\Lambda$ dependent terms in 
Eq.~\eqref{eq:defbssquare} which are only present in the gluon piece:
\begin{eqnarray}
\left(\frac{g_\mu^2 \mu^{-2\epsilon}N_c^2}{\pi(2\pi)^{D+1}} \right)^{-1}
\int\dktwo\int dy_2\,\bssquare{\qa}{\qb}{\ktwo}{\kone} \Big|_{s_\Lambda}
&&\nonumber\\
&&\hspace{-8cm} ~\equiv~
\int\dktwoeps \int dy_2 \, 
\frac{A^{\rm singular}_{\text{gluons}}}{\qa^2\qb^2}\, \theta(s_\Lambda-s_{12})
~=~ \sum_{i=I}^{IV} \mathcal{S}_i,
\label{eq:allJs}
\end{eqnarray}
where we have used the numbering $(I,II,III,IV)$ corresponding to, 
respectively, $({\rm Gluon|_{coll_1}},{\rm Gluon|_{coll_2}},
{\rm Gluon|_{soft}},{\rm Gluon|_{soft \& coll}})$ in 
Eq.~\eqref{eq:differentterms}.

To calculate each of the $\mathcal{S}_i$ terms we start by transforming the rapidity 
integral into an integral over $x$ in the form 
$\int d\Delta y=\int\frac{dx}{x(1-x)}$. We consider $s_{\Lambda}$ much larger 
than any of the typical transverse momenta. In the limit of large $s_\Lambda$ 
the theta function $\theta(s_\Lambda-\shat)$ amounts to the  
limits $\frac{\kone^2}{s_\Lambda}+\orderslambda{-2}$ and 
$1-\frac{\ktwo^2}{s_\Lambda}+\orderslambda{-2}$ for the $x$ integral. 

We firstly consider $\mathcal{S}_{III}$ which is 
\begin{eqnarray}
-\int\dktwoeps\int_{\frac{\kone^2}{s_\Lambda}}^{1-\frac{\ktwo^2}{s_\Lambda}}\frac{dx}{x(1-x)}\frac{1}{2\,\kone^2\ktwo^2} &=& \frac{-\pi}{(4\pi)^\epsilon}\frac{1}{\delt^2}\frac{\Gamma(1-\epsilon)\Gamma(\epsilon)^2}{\Gamma(2\epsilon)} 
\nonumber\\
&&\hspace{-6.5cm}\times\left(\ln\frac{s_\Lambda}{\delt^2}+\psi(1-\epsilon)-\psi(\epsilon)+\psi(2\epsilon)-\psi(1)\right)\left(\frac{\delt^2}{\mu^2}\right)^{\epsilon}+\orderslambda{-1}.
\label{eq:SIII}
\end{eqnarray}
We are only interested in the logarithmic dependence on $s_\Lambda$ and hence 
we do not need to calculate $\orderslambda{-1}$ or $s_\Lambda$ independent 
factors.

The next $s_\Lambda$ contribution we calculate is $\mathcal{S}_{IV}$ which reads
\begin{eqnarray}
\int\dktwoeps\int_{\frac{(\delt-\ktwo)^2}{s_\Lambda}}^{1-\frac{\ktwo^2}{s_\Lambda}}\frac{dx}{x(1-x)}
\left(\frac{(1-x)^2}{\ktwo^2(\ktwo-(1-x)\delt)^2}+\frac{x^2}{\ktwo^2(\ktwo-x\delt)^2}\right)&&\nonumber\\
&&\hspace{-12cm}= \int\dktwoeps\Bigg[\frac{2}{(\delt-\ktwo)^2\ktwo^2}\ln\frac{s_\Lambda}{\ktwo^2}+\frac{2\,(\delt-\ktwo)\ktwo}{(\delt-\ktwo)^2\ktwo^2\sqrt{\ktwo^2\delt^2-(\delt\ktwo)^2}}\nonumber\\
&&\hspace{-14.6cm}\phantom{2\int\dktwoeps\Bigg[}\times\left(
 \arctan\frac{\delt(\delt-\ktwo)}{\sqrt{\ktwo^2\delt^2-(\delt\ktwo)^2}}
+\arctan\frac{\delt\ktwo}{\sqrt{\ktwo^2\delt^2-(\delt\ktwo)^2}}\right)\Bigg]+\orderslambda{-1}.
\end{eqnarray}
The part with logarithmic $s_\Lambda$ dependence can be calculated analytically:
\begin{eqnarray}
\int\dktwoeps\frac{1}{(\delt-\ktwo)^2\ktwo^2}\ln\frac{s_\Lambda}{\ktwo^2} &=& 
\frac{\pi}{(4\pi)^\epsilon}\frac{1}{\delt^2}\frac{\Gamma(1-\epsilon)\Gamma(\epsilon)^2}{\Gamma(2\epsilon)}\nonumber\\
&&\hspace{-4cm}\times \left(\ln\frac{s_\Lambda}{\delt^2}+\psi(1-\epsilon)-\psi(\epsilon)+\psi(2\epsilon)-\psi(1)\right)\left(\frac{\delt^2}{\mu^2}\right)^{\epsilon}.
\end{eqnarray}
It is then clear that this logarithmic $s_\Lambda$ contribution cancels 
against that of $\mathcal{S}_{III}$ in Eq.~\eqref{eq:SIII}.

Let us proceed now to show that the contribution of $\mathcal{S}_{I}$ is directly of $\orderslambda{-1}$ and does not contribute with any logarithm of $s_\Lambda$. In the 
relevant integral we introduce the change of variable 
$\ktwo\to\lambd=(1-x)\delt-\ktwo$ and obtain
\begin{eqnarray}
\int\dlambdeps\int_{\frac{\lambd^2}{s_\Lambda}}^{1-\frac{\lambd^2}{s_\Lambda}}\frac{dx}{x(1-x)}
\left(\frac{x^2(1-x)^2}{\lambd^2(\lambd^2+x(1-x)\delt^2)}\right)&=&\nonumber\\
&&\hspace{-9.8cm} \int\dlambdeps
\left(\frac{1}{\delt^2\lambd^2 }-\frac{2\ln\left(1+\frac{\delt^2+\sqrt{\delt^2(\delt^2+4\lambd^2)}}{2\lambd^2}\right)}{\delt^2\sqrt{\delt^2(\delt^2+4\lambd^2)}}\right)+\orderslambda{-1}.
\end{eqnarray}
We do not write here the lengthier but similar expression which corresponds 
to $\mathcal{S}_{II}$ and also only contributes to $\orderslambda{-1}$.

With this we have shown that the sum of different terms
 in Eq.~\eqref{eq:allJs} 
is free of logarithmic dependences on $s_\Lambda$ proving, in this way, 
that the remaining $\orderslambda{-1}$ corrections vanish at large values 
of $s_\Lambda$. In particular, it is possible to take the 
$s_\Lambda \rightarrow \infty$ limit in order to completely eliminate the 
dependence on this scale. This is convenient in the fully inclusive 
case where it is very useful to write a Mellin transform in the $k_T$ 
dependence of the NLO BFKL kernel.

If we perform this $s_\Lambda \rightarrow \infty$ limit then  
$\mathcal{S}_{III}$ and $\mathcal{S}_{IV}$ can be put together and their sum is
\begin{align}
\mathcal{S}_{III}+\mathcal{S}_{IV} =& \int_0^1\frac{dx}{x(1-x)}\int\dktwoeps\left[\frac{1}{2\shat}\left(\frac{1-x}{x\ktwo^2}+\frac{x}{(1-x)\kone^2}\right)-\frac{1}{2\kone^2\ktwo^2}\right]\non
=& \int_0^1\frac{dx}{2x(1-x)}\int\dktwoeps\Bigg[\frac{(1-x)^2}{\ktwo^2(\ktwo-(1-x)\delt)^2}\non
&\quad\hspace{4cm}+\frac{x^2}{\kone^2(\kone-x\delt)^2}-\frac{1}{\ktwo^2(\delt-\ktwo)^2}\Bigg]\non
=&\frac{1}{\delt^2}\frac{\pi}{(4\pi)^\epsilon}\frac{\Gamma(1-\epsilon)\Gamma^2(1+\epsilon)}{\epsilon \, \Gamma(1+2\epsilon)}\left(\frac{1}{\epsilon}+2\psi(1)-2\psi(1+2\epsilon)\right)\left(\frac{\delt^2}{\mu^2}\right)^\epsilon.\label{eq:sthreeplusfour}
\end{align}
Regarding $\mathcal{S}_{I}$ the integration gives us
\begin{eqnarray}
\mathcal{S}_{I}&=&-2\int_{0}^{1}\frac{dx}{x(1-x)}\int\dlambdeps\left[\frac{x^2(1-x)^2}{\lambd^2(\lambd^2+x(1-x)\delt^2)}\right]\nonumber\\
&=& -2 \int_{0}^{1}\frac{dx}{x(1-x)}\left[\frac{\pi}{(4\pi)^\epsilon}\frac{x(1-x)}{\delt^2}\frac{\Gamma(1-\epsilon)\Gamma(\epsilon)}{\Gamma(1+\epsilon)}\left(\frac{x(1-x)\delt^2}{\mu^2}\right)^\epsilon\right] \nonumber\\
&=& -\frac{2}{\delt^2}\frac{\pi}{(4\pi)^\epsilon}\frac{\Gamma(1-\epsilon)\Gamma(1+\epsilon)^2}{\epsilon\Gamma(2+2\epsilon)}\left(\frac{\delt^2}{\mu^2}\right)^\epsilon.
\end{eqnarray}
The contribution from $\mathcal{S}_{II}$ is more complicated and the relevant integral 
can be obtained in the following way:
\begin{eqnarray}
\int\dlambdeps\frac{E^2}{8\qa^2\qb^2\shat^2}
&=& \int\dlambdeps\frac{x^2(1-x)^2E^2}{8\qa^2\qb^2\lambd^4} \nonumber\\
&&\hspace{-4cm}= \int\dlambdeps \bigg[\frac{x^2(1-x)^2(2x-1)^2\delt^2}{2\qb^2\lambd^2 (x(1-x)\delt^2+\lambd^2)}\non
&&\hspace{-3.5cm}-\frac{x^3(1-x)^3(2x-1)^2\delt^2\qa^2}{\qb^2\lambd^4(x(1-x)\delt^2+\lambd^2)}+\frac{x^4(1-x)^4(2x-1)^2\delt^4\qa^2}{2\qb^2\lambd^4(x(1-x)\delt^2+\lambd^2)^2}\non
&&\hspace{-3.5cm}-\frac{4x^3(1-x)^3 (\delt\lambd)(\lambd\qa)}{\qb^2\lambd^2(x(1-x)\delt^2+\lambd^2)}+\frac{2x^4(1-x)^4(\delt\lambd)^2\qa^2}{\qb^2\lambd^2(x(1-x)(\delt^2+\lambd^2)^2}\bigg]\\
&&\hspace{-4cm}= \frac{\pi}{(4\pi)^\epsilon}\frac{\Gamma(2-\epsilon)\Gamma(\epsilon)}{\Gamma(1+\epsilon)}\frac{\left(x(1-x)\delt^2\right)^{\epsilon-1}}{\mu^{2\epsilon}}\bigg[\frac{1}{1-\epsilon}\frac{x^2(1-x)^2(2x-1)^2\delt^2}{\qb^2}\non
&&\hspace{-3.5cm}+\frac{1}{1-\epsilon}\frac{x^2(1-x)^2(2x-1)^2\qa^2}{\qb^2}-\frac{2-\epsilon}{1-\epsilon}\frac{x^2(1-x)^2(2x-1)^2\qa^2}{2\qb^2}\non
&&\hspace{-3.5cm}-\frac{2}{1-\epsilon^2}\frac{x^3(1-x)^3\delt\qa}{\qb^2}+\frac{1}{1+\epsilon}\frac{x^3(1-x)^3\qa^2}{\qb^2}\bigg]
\end{eqnarray}
We now need to integrate it over $x$ to obtain:
\begin{equation}
  \mathcal{S}_{II} = \frac{1}{\delt^2}\frac{\pi}{(4\pi)^\epsilon}\frac{\pi(1+\epsilon)\Gamma(2+\epsilon)}{\sin(\pi\epsilon)\Gamma(4+2\epsilon)}\left(\frac{\delt^2}{\mu^2}\right)^\epsilon .
\end{equation}
 This result gives the same poles in $\epsilon$ as the result given in \cite{Ostrovsky:1999kj}, but differs for the finite contribution.
To obtain all the $\epsilon$ poles we now also include the quark contributions 
present in Eq.~\eqref{eq:defbssquare}. We denote them as
\begin{eqnarray}
\int\dktwoeps \int dy_2 \, \frac{A^{\rm singular}_{\text{quarks}}}{\qa^2\qb^2} &=& 
\sum_{i=V}^{VI} \mathcal{S}_i,
\label{eq:onlyquarksinBs}
\end{eqnarray}
where the correspondence with Eq.~\eqref{eq:differentterms} is 
$(V,VI) \rightarrow ({\rm Quark|_{coll_1}},{\rm Quark|_{coll_2}})$. 
Adding everything up the sum of all the terms reads
\begin{equation}
  \sum_{i=I}^{VI}\mathcal{S}_i = \frac{1}{\delt^2}\frac{\pi\Gamma(1-\epsilon)}{(4\pi)^\epsilon}\left(\frac{\delt^2}{\mu^2}\right)^\epsilon\left[\frac{1}{\epsilon^2}-\frac{\beta_0}{2N_c}\frac{1}{\epsilon}+\frac{67}{18}-\frac{5n_f}{9N_c}-\frac{5\pi^2}{6}+\order{\epsilon}\right].
\end{equation}
The final expression for Eq.~\eqref{eq:defbssquare} is then
\begin{multline}
  \int\dktwo\int dy_2 \bssquare{\qa}{\qb}{\kjet-\ktwo}{\ktwo} = \\
\frac{\bar{g}_\mu^4 \mu^{-2\epsilon}}{\pi^{1+\epsilon}\Gamma(1-\epsilon)}\frac{4}{\kjet^2}\left(\frac{\kjet^2}{\mu^2}\right)^\epsilon\left[\frac{2}{\epsilon^2}-\frac{\beta_0}{N_c}\frac{1}{\epsilon}+\frac{67}{9}-\frac{10n_f}{9N_c}-\frac{5\pi^2}{3}+\order{\epsilon}\right] \label{eq:resultbsquare}.
\end{multline}
When we combine this result with the singular terms of Eq.~\eqref{eq:kernelv} 
then we explicitly prove the cancellation of any singularity in our 
subtraction procedure to introduce the jet definition. The finite remainder 
reads
\begin{eqnarray}
\frac{\bar\alpha_s^2(\mu^2)}{\pi}
\frac{1}{\kjet^2}
\left[-\frac{\beta_0}{4 N_c}\ln{\frac{\kjet^2}{\mu^2}}+\frac{1}{12}\left(
4 - 2 \pi^2 + 5 \frac{\beta_0}{N_c}\right)\right].
\end{eqnarray}
We have already discussed the logarithmic term due to the running of the 
coupling in Eq.~\eqref{eq:runningcoupling}. The non--logarithmic part is 
similar to that present in other calculations involving soft gluon 
resummations~\cite{GBesquema} where terms of the form 
\begin{eqnarray}
{\bar \alpha}_s \left(1+ {\cal S} \,{\bar \alpha}_s\right)
\end{eqnarray}
appear and offer the possibility to change from the $\overline{\rm MS}$ 
renormalization scheme to the so--called {\sl gluon--bremsstrahlung} (GB) 
scheme by shifting the position of the Landau pole, {\it i.e.}
\begin{eqnarray}
\Lambda_{\rm GB} &=& \Lambda_{\rm \overline{\rm MS}} 
\exp{\left({\cal S}\frac{2 N_c}{\beta_0}\right)}.
\end{eqnarray}
The factor $S$ differs from ours in the $\pi^2$ term:
\begin{eqnarray}
{\cal S} &=& \frac{1}{12}\left(4 - \pi^2 + 5 \frac{\beta_0}{N_c}\right).
\end{eqnarray}
The origin of this discrepancy lies in the fact that we used the simplest 
form of subtraction procedure. In the Appendix we suggest a different 
subtraction term which is more complicated in the sense that it 
substracts a larger portion of the matrix element in addition to the infrared 
divergent pieces. When this is done and we put together the divergent pieces 
of Eq.~\eqref{eq:kernelv} and the second line of 
Eq.~\eqref{eq:resultbsquarenew} then we recover the same ${\cal S}$ term.

\section{Conclusions}

In this paper we have extended the NLO BFKL calculations to derive a NLO jet
production vertex in $k_T$--factorization. Our procedure was to 
`deconstruct' the NLO BFKL kernel to introduce a jet definition at NLO in a 
consistent way. After a 
careful study of the different energy scales and contributions to the kernel 
we were able to show the infrared finiteness of this jet vertex and its 
dependence on the scale $s_\Lambda$, which separates MRK from QMRK. As the 
central result of this
paper, we have defined the jet production vertex \eqref{eq:freeofsing} in
terms of longitudinal momentum fractions, and we have explicitly given the necessary
subtraction, both at the matrix element level \eqref{eq:differentterms} as
well as integrated over the corresponding phase space \eqref{eq:resultbsquare}. 
Our calculations also suggest that the natural scale for the running 
of the coupling at the jet vertex is the square of the 
transverse momentum of the jet \eqref{eq:runningcoupling}.
We have shown how this vertex can be used in the context of $\gamma^*\gamma^*$ 
or hadron--hadron scattering \eqref{ppfinal} to calculate inclusive single jet cross sections. 
For this purpose we have formulated, on the basis of the NLO BFKL equation, a
NLO unintegrated gluon density valid in the small--$x$ regime.

In our analysis we have been careful to retain the dependence upon the 
energy scale $s_\Lambda$ which appears at NLO accuracy and separates 
multi--Regge kinematics from quasi--multi--Regge kinematics. In the NLO 
calculation of the total cross section, one may be tempted to take the limit 
$s_\Lambda \to \infty$, thus disregarding the $1/s_\Lambda$ corrections to the 
NLO BFKL kernel. However, when discussing inclusive (multi-) jet production 
one has to remember that $s_\Lambda$ has a concrete physical meaning: it 
denotes the lower cutoff of rapidity gaps and thus directly enters the 
rapidity distribution of multi--jet final states. In a self--consistent 
description then also the evolution of the unintegrated gluon density has to 
depend upon this scale. 

Hence we are well prepared for our next step, the numerical study of single or
multiple jet production in hadron--hadron collisions at the LHC. One issue to 
be covered will be the question of handling the running of coupling. Further 
applications of our NLO $k_T$--formalism include $W$ and $Z$ as well as
heavy flavor production in the small--$x$ region. Compared to the results 
presented in this paper, these applications require the calculation of 
further production vertices; however, for the treatment of the different 
scales and of the unintegrated gluon density all basic ingredients have been 
collected in this paper.\\ \\

\noindent
{\bf Acknowledgements:} A.S.V. thanks the Alexander--von--Humboldt
Foundation for financial support. F.S. is supported by the Graduiertenkolleg 
``Zuk\"unftige Entwicklungen in der Teilchenphysik''. Helpful discussions with 
V.~S.~Fadin and L.~N.~Lipatov are gratefully acknowledged.

\begin{appendix}

\section{Alternative subtraction term}

In this Appendix we present an alternative subtraction term which does not 
make use of the simplifications $A_{(3)}+A_{(4)}\to 2A_{(4)}$ and 
$A_{(5)}+A_{(6)}\to 2A_{(5)}$ which we used in Eqs.~(\ref{eq:agluonssoft}, \ref{eq:agluonssoftcoll2}). These limits are valid in the kinematic regions leading to IR--divergences and hence they do provide the correct $\epsilon$ poles. However, they also alter the finite terms. Here we want to study also this finite 
part as accurately as possible and hence we do not take these limits  
but use the complete sum
\begin{equation}
  \label{eq:alternativebase}
  A_{(1)}+A_{(2)}+A_{(3)}+A_{(4)}+A_{(5)}+A_{(6)}\;+\;A_{\rm MRK} 
\end{equation}
as the gluonic subtraction term.

The full gluonic matrix element written in Eq.~\eqref{eq:Agluons} contains 
spurious UV--divergences which are cancelled when combined with the MRK contribution. One fourth of the MRK contribution cancels the UV--divergence of $A_{(4)}$ while another fourth cancels that of $A_{(6)}$. The remaining half cancels the UV--divergence of two terms present in Eq.~\eqref{eq:Agluons}:
\begin{align}
 A_{(7)}\equiv& -\frac{\qa^2\qb^2}{4}\left(\frac{1-x}{x}\frac{1}{\ktwo^2\that}+\frac{x}{1-x}\frac{1}{\kone^2\uhat}\right)\\
 A_{(8)} \equiv& \frac{\qa^2\qb^2}{4\Sigma}\left(\frac{1-x}{x}\frac{1}{\ktwo^2}+\frac{x}{1-x}\frac{1}{\kone^2}\right),
\end{align}
which are IR--finite and hence so far not included in the subtraction term.

By doubling $A_{(4)}$ and $A_{(5)}$ in the subtraction term constructed in Eq.~\eqref{eq:differentterms} also their spurious UV--divergences are doubled and thus completely cancelled by the MRK contribution. But Eq.~\eqref{eq:alternativebase} so far only contains half of the spurious UV--divergences of the full matrix element in such a way that half of the MRK contribution is not compensated. Therefore a subtraction term based on Eq.~\eqref{eq:alternativebase} which is also free from spurious UV--divergences should also include $A_{(7)}$ and $A_{(8)}$ and reads
\begin{eqnarray}
  \widetilde{A}_{\rm gluons}^{\rm singular} &=& A_{(1)}+A_{(2)}+A_{(3)}+A_{(4)}+A_{(5)}+A_{(6)}+A_{\rm MRK}+A_{(7)}+A_{(8)} \nonumber\\
&&\hspace{-1.5cm}= A_{(1)}+A_{(2)}+A_{(3)}+\big(A_{(5)}-A_{(4)}\big)+A_{(6)}+\frac{A_{\rm MRK}}{2}+A_{(7)}+A_{(8)}.
\end{eqnarray}
If we now define $\mathcal{S}_{(3,6,7,8)}$ and $\mathcal{S}_{\rm MRK}$ as we 
did in Eq.~\eqref{eq:allJs} we get a new integrated subtraction term from 
the previous Eq.~\eqref{eq:resultbsquare} by replacing
\begin{equation}
\mathcal{S}_{III}+\mathcal{S}_{IV} = \frac{1}{\delt^2}\frac{\pi\Gamma(1-\epsilon)}{(4\pi)^\epsilon}\left(\frac{\delt^2}{\mu^2}\right)^\epsilon\left[\frac{1}{\epsilon^2}-\frac{5\pi^2}{6}+\order{\epsilon}\right]
\end{equation}
with
\begin{equation}
 \frac{1}{2}\left(\mathcal{S}_{III}+\mathcal{S}_{IV}\right)+\mathcal{S}_{(3)}+\mathcal{S}_{(6)}+\frac{\mathcal{S}_{\rm MRK}}{2}+\mathcal{S}_{(7)}+\mathcal{S}_{(8)} .
\end{equation}
The results for $\mathcal{S}_{(3)}$ and $\mathcal{S}_{(6)}$ can be easily obtained from Eqs.~(C.43) and (C.40) of Ref.~\cite{Kotsky:1998ug}:
\begin{eqnarray}
  \mathcal{S}_{(3)} &=& \frac{1}{\delt^2}\frac{\pi\Gamma(1-\epsilon)}{(4\pi)^{\epsilon}}\left(\frac{\delt^2}{\mu^2}\right)^\epsilon\Bigg[\frac{1}{2\epsilon^2}+\frac{1}{2\epsilon}\ln\frac{\qa^2\qb^2}{\delt^4}-\frac{\pi^2}{12}+\frac{1}{4}\ln^2\frac{\qa^2}{\qb^2}\nonumber\\
&+&\frac{\qa^2\qb^2\big(\delt(\qa-\qb)\big)}{\delt^2(\qa-\qb)^2}\Bigg\{\frac{1}{2}\ln\left(\frac{\qa^2}{\qb^2}\right)\ln\left(\frac{\qa^2\qb^2\delt^4}{(\qa^2+\qb^2)^4}\right)\nonumber\\
&-&\text{Li}_2\left(-\frac{\qa^2}{\qb^2}\right)+\text{Li}_2\left(-\frac{\qb^2}{\qa^2}\right)\Bigg\}
-\frac{\qa^2\qb^2}{2}\left(1-\frac{\big(\delt(\qa-\qb)\big)^2}{\delt^2(\qa-\qb)^2}\right)\nonumber\\
&&\hspace{2cm}\times\left(\int_0^1-\int_1^\infty\right)dz\frac{\ln\left(\frac{(z\qa)^2}{\qb^2}\right)}{(\qb+z \qa)^2}+\order{\epsilon}\Bigg],\\
\mathcal{S}_{(6)} &=& \frac{1}{\delt^2}\frac{\pi\Gamma(1-\epsilon)}{(4\pi)^{\epsilon}}\left(\frac{\delt^2}{\mu^2}\right)^\epsilon\left[\frac{1}{\epsilon^2}-\frac{\pi^2}{6}+\order{\epsilon}\right].
\end{eqnarray}
Due to the UV--singularity of $A_{\rm MRK}$ we regularize the $x$ integration by a cutoff $\delta$ to obtain
\begin{eqnarray}
  \mathcal{S}_{\rm MRK} &=&   -\int_\delta^{1-\delta} \frac{dx}{x(1-x)}\int\frac{\dktwo}{\mu^{2\epsilon}(2\pi)^{D-4}}  \frac{1}{\ktwo^2(\delt-\ktwo)^2}\nonumber\\
&=&\frac{1}{\delt^2}\frac{\pi\Gamma(1-\epsilon)}{(4\pi)^{\epsilon}}\left(\frac{\delt^2}{\mu^2}\right)^\epsilon\frac{\Gamma^2(\epsilon)}{\Gamma(2\epsilon)}2\ln\frac{\delta}{1-\delta}.
\end{eqnarray}
Making use of $2\qa\kone-\qa^2=\that+\kone^2/x$ we can decompose Eq.~(C.41) of Ref.~\cite{Kotsky:1998ug} into one integration very similar to that of $\mathcal{S}_{\rm MRK}$ and another one which can be transformed to give $\mathcal{S}_{(7)}$.
\begin{align}
\mathcal{S}_{(7)} =&  \frac{1}{\delt^2}\frac{\pi\Gamma(1-\epsilon)}{(4\pi)^{\epsilon}}\left(\frac{\delt^2}{\mu^2}\right)^\epsilon\Bigg[-\frac{1}{2}\frac{\Gamma^2(\epsilon)}{\Gamma(2\epsilon)}\ln\frac{\delta}{1-\delta}-\frac{1}{2\epsilon^2}-\frac{1}{2\epsilon}\ln\frac{\qa^2\qb^2}{\delt^4}\non
&\hphantom{\frac{1}{\delt^2}\frac{\pi\Gamma(1-\epsilon)}{(4\pi)^{\epsilon}}\left(\frac{\delt^2}{\mu^2}\right)^\epsilon\Bigg[}-\frac{1}{4}\ln^2\frac{\qa^2}{\qb^2}+\frac{\pi^2}{12}+\order{\epsilon}\Bigg].\\
\intertext{The two parts forming $A_{(8)}$ can be obtained from each other by 
the exchange $k_1\leftrightarrow k_2$ and we only need to double the 
calculation of one:}
   \mathcal{S}_{(8)} =&  2\int_\delta^{1-\delta} \frac{dx}{x(1-x)} \int\frac{\dkone}{\mu^{2\epsilon}(2\pi)^{D-4}}\frac{x}{4(1-x)}\frac{1}{\Sigma\kone^2}\\
=&2\int_\delta^{1-\delta} \frac{dx}{x(1-x)} \int\frac{\dkone}{\mu^{2\epsilon}(2\pi)^{D-4}}\non
&\times\frac{1}{4}\int_0^1 d\xi \frac{x^2}{\Big\{[\kone-\xi x\delt]^2+\xi(1-\xi)x^2\delt^2+\xi x(1-x)\delt^2\Big\}^2}\\
=&\frac{1}{2} \frac{\pi\Gamma(1-\epsilon)}{(4\pi)^\epsilon}\frac{1}{\delt^2}\left(\frac{\delt^2}{\mu^2}\right)^\epsilon\int_\delta^{1-\delta} dx\frac{1}{1-x}B_{x}(\epsilon,\epsilon)\\
 =& \frac{1}{\delt^2}\frac{\pi\Gamma(1-\epsilon)}{(4\pi)^{\epsilon}}\left(\frac{\delt^2}{\mu^2}\right)^\epsilon\left[-\frac{1}{2}\frac{\Gamma^2(\epsilon)}{\Gamma(2\epsilon)}\ln\delta-\frac{1}{2\epsilon^2}-\frac{\pi^2}{12}+\order{\epsilon}\right] .
\end{align}
When we add up these new contributions the spurious UV--divergences indeed cancel and we can safely take the $\delta\to 0$ limit. Furthermore, the new subtraction term has the same pole structure and only different finite parts when compared to that in Eq.~\eqref{eq:differentterms} and its integrated form in Eq.~\eqref{eq:resultbsquare}. To complete the calculation we combine it with the 
corresponding unmodified quark part and obtain
\begin{eqnarray}
\int\dktwo\int dy_2 \btssquare{\qa}{\qb}{\kjet-\ktwo}{\ktwo} &=& \nonumber\\
&&\hspace{-7.5cm}\frac{\bar{g}_\mu^4 \mu^{-2\epsilon}}{\pi^{1+\epsilon}\Gamma(1-\epsilon)}\frac{4}{\kjet^2}\left(\frac{\kjet^2}{\mu^2}\right)^\epsilon 
\Bigg\{\frac{2}{\epsilon^2}-\frac{\beta_0}{N_c}\frac{1}{\epsilon}+\frac{67}{9}-\frac{10n_f}{9N_c}-\frac{4\pi^2}{3}\nonumber\\
&&\hspace{-7cm}+\frac{2\qa^2\qb^2\big(\delt(\qa-\qb)\big)}{\delt^2(\qa-\qb)^2}\Bigg[\frac{1}{2}\ln\left(\frac{\qa^2}{\qb^2}\right)\ln\left(\frac{\qa^2\qb^2\delt^4}{(\qa^2+\qb^2)^4}\right)\nonumber\\
&&\hspace{-2cm}-\text{Li}_2\left(-\frac{\qa^2}{\qb^2}\right)+\text{Li}_2\left(-\frac{\qb^2}{\qa^2}\right)\Bigg]\nonumber\\
&&\hspace{-7.5cm}
-\qa^2\qb^2\left(1-\frac{\big(\delt(\qa-\qb)\big)^2}{\delt^2(\qa-\qb)^2}\right)
\left(\int_0^1-\int_1^\infty\right)dz\frac{\ln\left(\frac{(z\qa)^2}{\qb^2}\right)}{(\qb+z \qa)^2}+\order{\epsilon}\Bigg\}
\label{eq:resultbsquarenew}.
\end{eqnarray}

\end{appendix}

\end{document}